\documentclass[10pt]{article}

\usepackage{numprint}
\usepackage{bbm}
\usepackage{amsmath}
\usepackage{xcolor}
\usepackage{graphicx}
\usepackage{xspace}
\usepackage{fancyvrb}
\usepackage{booktabs}
\usepackage{hyperref}

\usepackage[multiple]{footmisc}

\usepackage{blindtext}
\usepackage{geometry}
\begin{document}

\title{\Large \textbf{An Evaluation of Researchers' Migration Patterns in Europe using Digital Trace Data}}

\author{Jacopo Ghirri\footnote{These authors contributed equally to this work.}, Marta Mastropietro\footnotemark[1],\\ \text{Simone Vantini}, \text{Francesca Ieva}, \\
\textit{\large{Politecnico di Milano}}\\
\text{Matteo Fontana}\\
\textit{\large{Joint Research Centre, Ispra}}}

\date{}

\maketitle


\abstract{
The comprehension of the mechanisms behind the mobility of skilled workers’ is of paramount importance for policy making. The lacking nature of official measurements motivates the use of digital trace data extracted from ORCID public records. We use such data to investigate European regions, studied at NUTS2 level, over the time horizon of 2009 to 2020. We present a novel perspective where regions’ roles are dictated by the overall activity of the research community, contradicting the common brain drain interpretation of the phenomenon. We find that a high mobility is usually correlated with strong university prestige, high magnitude of investments and an overall good schooling level in a region.
}

\section{Introduction}

Ideas, technology and knowledge are assets of ever growing importance for every economy and, as the people who are able to offer such assets migrate, they can shift the balances in the race for innovation with unpredictable impact, especially on the long term. 
The global distribution of capabilities is indeed highly unequal \cite{yilmaz_human_2009} and a better understanding of what guides skills movements, and the movement of skilled workers can be a competitive advantage for determining the future leading countries in research and innovation.

Researchers' mobility in particular has been studied with exceptional care. Many studies have investigated its socioeconomic implications, such as the relationship between academic mobility and innovation \cite{han_will_2015,jonkers_research_2013,heitor_developing_2014} and analyses have been carried out on the effects of talents' loss on regions \cite{beine_brain_2001, lodigiani_revisiting_2016, yu_brain_2021}.

The interest is however not only academic, but it has serious policy implications. Policy makers are actively seeking ways to improve their regions' research activity and skill attraction capabilities, as testified by the numbers of already existing EU-sponsored initiatives \cite{european_committee_of_the_regions_addressing_2018} as well as by the 2022 State of the Union Address speech, where Ursula Von der Leyen, the European Commission president, made clear the importance of understanding skill's mobility and how to act on it: "We need much more focus in our investment on professional education and upskilling. [...] But we also have to attract the right skills to our continent, skills that help companies and strengthen Europe’s growth. [...] This is why I am proposing to make 2023 the European Year of Skills." \cite{UVdL_speech}.

Since the times of great outflows of scientists and technologists from Europe toward North America in the 1950s and early 1960s, the focus of the great majority of economists and sociologists has been the study of this phenomenon under the lens of brain drain, intended as a net migration imbalance between different regions \cite{lodigiani_revisiting_2016, cervantes_brain_2002}. In the global network of researchers' mobility it was commonly believed that there are receiving regions and provider regions, where the former are able to attract skills, hence receiving the consequent benefits, while the latter are losing talents and are less competitive in innovation and research.

Such an interpretation is so widespread that it is often used interchangeably for indicating the migration phenomenon itself, however it has been recently studied as a part of a much wider mechanism that can generate benefits for all regions involved, if handled correctly, and not only for the immediate downside that a loss of talent can cause. Some studies explicitly distinguish between a "brain effect" and a "drain effect" \cite{beine_brain_2001} and there is an even wider interest in the impact that a brain drain phenomenon could have on different aspects of innovation \cite{siekierski_international_2018}, with many findings being discordant with the idea of researcher's migrations being strictly beneficial for some regions and detrimental for others.
On a parallel track new ideas, like brain circulation \cite{yu_brain_2021}, are becoming increasingly present in socioeconomic literature, challenging the historically consolidated belief of researchers' mobility being characterized by a net brain drain and brain gain.

Empirical exploration of this phenomenon is particularly challenging, as data about migration flows tend to be inconsistent across countries and are usually not timely neither provided with the required spatial granularity. It is moreover exceptionally rare to have such data being associated with further information about the age, occupation or other social characteristics of migrants. Such issues arise from the fact that "the popularity and relevance of migration has outpaced substantial improvements in the systematic measurement of migration" \cite{tjaden_measuring_2021} or, in some cases, by the straight-out nonexistence of data relevant for the phenomenon to be studied.\cite{zagheni_inferring_2014}.

The use of unconventional data sources for addressing research mobility, in particular bibliometric data \cite{bohannon_introducing_2017,Boudry_use_2020} and digital trace data \cite{JRC_digital_trace}, has recently been subject of more and more investigation, with ever-growing interest in their application \cite{miranda-gonzalez_scholarly_2020,urbinati_measuring_2021}.
We follow this stream of research by proposing the use of ORCID records as an attempt to obtain a reliable, traceable and meaningful sample of the overall researcher's population, on which a statistical analysis of the migration phenomenon can be conducted. In particular we focus on EU27 plus United Kingdom, Norway, Iceland and Switzerland, treated at the regional NUTS2 level \cite{nuts_definition}, over the time horizon spanning from 2009 to 2020.

Our work is carried out by extracting and exploiting a new data collection of ORCID profiles, on which it is possible to conduct data driven analyses. Strong of this we aim at comprehending whether we can talk about brain drain in Europe or if there are other perspectives for understanding the researchers' migration phenomenon. Given an answer to this question, our final purpose is to select regional characteristics able to correlate with such migrations and exploring how a policy maker could theoretically improve a region's research activity.

With this aim in mind, Section \ref{section:Data and preprocessing} illustrates our procedure for extracting the data used through our analysis from ORCID affiliations, it examines the selection of the time-space horizon for the analysis as a mean to retain representativeness of our sample and it presents the regional characteristics we select. Section \ref{section: Preliminary Analysis} contains an exploratory analysis on our data which covers both time and space dimensions of our problem. Section \ref{section: Mobility Analysis} aims at providing a meaningful characterization of the migration phenomenon, using amongst other instruments an ad-hoc adjusted ANOVA test detailed in Section \ref{section: ANOVA NPC}. Finally Section \ref{section: Models} explores the correlation between the phenomenon and the regional characteristics.

\subsection{A regional perspective}

The topic of skills migration and brain drain in particular is usually framed under a national perspective, but it is not rare to find discussions on researchers diaspora in the boundaries of the same country, especially in the context of long-standing regional socioeconomic inequalities. Notable are the examples of Italy \cite{asso2021346} and Germany \cite{kaplan201661}, historically associated with strong internal movements of high and low-skilled workers.

We hence analyze the phenomenon of researchers' migrations with a specific focus on regional borders, as we believe that this perspective, apart from being able to capture such granular migrations invisible at a higher level, would allow for exploring consistency or discrepancies with respect to the national trends.

For these reasons we will present all the analyses at regional NUTS2 level \cite{nuts_definition}, with a consistent and parallel check on the results obtained at the national level, reported in Appendix \ref{Appendix C: National Analysis}.


\section{Data and Pre-Processing}
\label{section:Data and preprocessing}

This Section covers our procedure for extracting the data we use over the course of this analysis, it addresses the major issues that could be raised against it and it explains the choices that were made to address them. Section \ref{subsection: ORCID data} focuses on the migration data extracted from ORCID, while the focus of Section \ref{subsection:coovariates} are regional characteristics.

\subsection{Migration Dataset}
\label{subsection: ORCID data}

Since official data for tackling the phenomenon of researchers migration is unavailable, we extract the needed migration patterns from ORCID public data. ORCID (Open Researcher and Contributor ID) is a non profit organisation developed in 2012 for all people involved in research, scholarship, and innovation, which provides a unique identifier to each enrolled member \cite{aboutorcid}.

Interest in digital trace as a way for solving data scarcity on migration has recently grown \cite{alonso_mapping_2022,bosco_data_2022} and we base our approach on ORCID due the reliability of its ID definition, which is one of the most attractive features with respect to other widely adopted bibliometric sources like Scopus \cite{scopus}: Scopus author ID lacks the level of precision offered by ORCID and it suffers from having authors with the same, or similar, names being grouped under the same identifier \cite{miranda-gonzalez_scholarly_2020}, an issue that ORCID aims at solving \cite{butler_scientists_2012}.

ORCID public data \cite{Blackburn_ORCIDDATA2021} is a collection of the publicly available information associated to the ORCID record of each user \cite{orcid_policy}, it contains information on their works, publications and affiliations. From its public database we are able to extract and trace migration patterns of registered users down to the municipality levels, across 31 European countries (EU27 plus United Kingdom, Norway, Iceland and Switzerland) in the time horizon spanning from 2009 to 2020. We decide to explore the phenomenon from a regional perspective, utilizing the NUTS2 2021 classification for territorial units \cite{nuts_definition}. 
The only exception we introduce with respect to the official definitions is London: we need to treat it like a single NUTS2, which we label as \textit{UKI0}, since the level of granularity required by the sub-regions of the city is not obtainable from municipalities in ORCID extracted data.

\subsubsection{Representativeness}
The core assumption of our analysis is that the sample of researchers we gather from ORCID's users is a representative sample of the global researchers' population. Such a premise is not easy to verify and has been object of studies in literature. One of the principal critiques raised in \cite{youtie_tracking_2017} is the uneven adoption of ORCID at national level, pointing out an under representation of Russia and Asian nations with respect to Europe. Authors of \cite{bohannon_introducing_2017} confirm a continental bias as well, on top of a time bias originated from the quicker adoption of ORCID from the younger portion of the researchers' population, being ORCID active since 2012.

\begin{figure}
\centering
\caption{\label{figure: nuts2 map}Main spatial dominion of the investigation.}
\makebox{\includegraphics[scale=0.4]{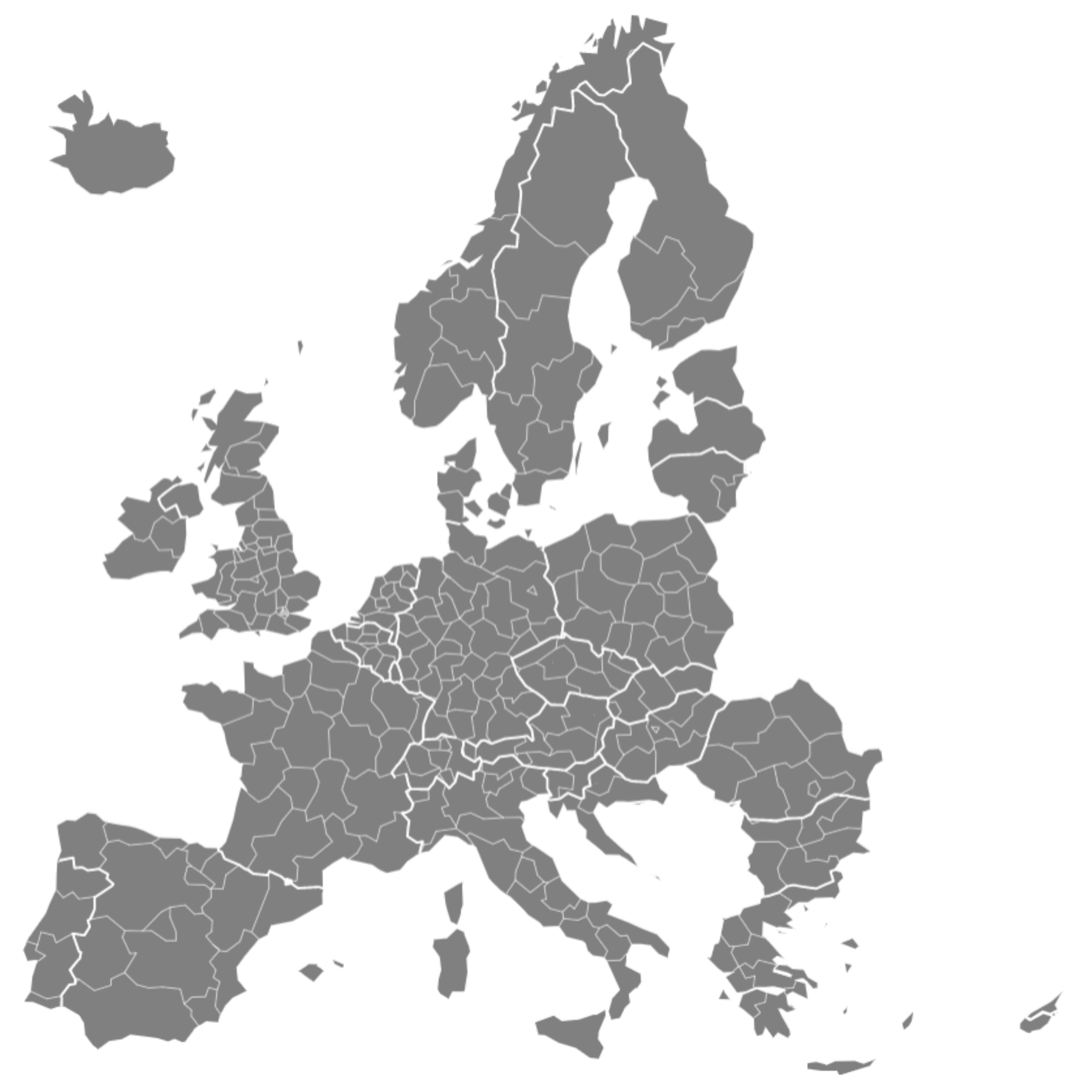}}
\end{figure}

We try to address and treat such issues by means of apposite restrictions of our analysis.
Firstly, we restrict the time horizon and we focus on the migrations that took place between Jan 1, 2009 and Dec 31, 2020. We have no straightforward way to verify the age distribution of ORCID users, hence no way to quantify the disproportion towards younger researchers, but our assumption is that by focusing on a fairly recent time horizon such age imbalances would not be particularly significant.\\
We treat the geographical imbalance by focusing the study on the regions of 31 European countries: EU27, United Kingdom, Norway, Iceland and Switzerland. Indeed, on top of Europe having a stronger adoption of ORCID, between the European Union freedom of travel and Schengen Area regulations we can assume migrations not to be significantly affected by bureaucracy and international relationships between countries.
We hypothesize the same holds for UK despite the happening of Brexit, since our analysis' time span ends in 2020 and United Kingdom's travel regulations were in a transition period up to December 31, 2020. A visualization of the main selected area of interest for the analysis is reported in Figure \ref{figure: nuts2 map}.\\ 
Peripheral regions, namely Svalbard Islands, Ceuta, Melilla, Canarias, Guadeloupe, Guyane, La Réunion, Mayotte, Martinique, Malta, autonomous regions of Açores and Madeira are not reported in our maps for the sake of better visualization on the principal regions, but are nevertheless considered in the analysis. 

\subsubsection{Data Extraction}
\label{subsub: ORCID Data extraction}
In the scope of our analysis the most important feature of ORCID public data is the presence of affiliations reported by each ORCID user. Amongst them we can find, where reported, information on the municipality where the affiliation took place, its beginning and its end dates.

The first step in our procedure requires to map the reported municipalities into the regional NUTS2 definition, we do so by means of the table available at \cite{geonames}, which allows us to map municipality names into postal codes, and those available at \cite{postcodenuts}, which map postal codes into NUTS regions. Since municipalities on ORCID affiliations were manually inserted by users, and sometimes are missing, we intervened to correct spelling errors (e.g. "Milnao" in place of "Milano"), language mismatches (e.g. "Milan" in place of "Milano") and similar issues. We deem the the results satisfactory when approximately 95\% of the affiliations that took place in each given country are successfully mapped into the respective NUTS2 region.

Once the affiliations are mapped it is sufficient to order them, for each ORCID user, with respect to time. We can then infer, based on the end year of one affiliation and the start year of the next, when a migration has occurred. This is done if two consecutive affiliations happened to be in two different NUTS2 regions, therefore capturing local migrations as well as international ones.
We preferably use the start year of the new affiliation as year of migration, when it is not reported we use the end year of the previous one. Repeating this procedure for every ORCID user we get a new table, indicating for every year and for every couple of regions how many researchers ended an affiliation in the first to start a new one in the second.
It is then sufficient to filter such table with respect to the desired time-space horizon in order to get the final dataset which constitutes the cornerstone of this analysis.

During the course of the analysis we will focus on two empirical settings. When we study the total number of people entering or leaving a region of interest, we are counting all ORCID users, even those coming from or going to regions outside of our spatial domain, as the focus is on the specific region in question and the origin or destination of a migration bear no impact. However, when we study the phenomenon within a network framework, we restrict the data exclusively to the migrations which are internal to the spatial dominion itself, as to obtain a coherent definition of network with well defined nodes.

\subsection{Regional Characterization}
\label{subsection:coovariates}

Part of the objective of this analysis is to study the phenomenon under the lens of regional-level covariates. For this reason we collect a selection of variables we deem to be tied with the phenomenon and interesting from a policy standpoint.
Each Section focuses on one particular variable, illustrating how we define, extract and interpret it. Together with standard pre processing to obtain the needed tables for each variable, we exploit the Amelia method as implemented in the homonymous R package \cite{honaker_amelia_2011} to impute unavailable data. This method is based on a multiple imputation model allowing smooth time trends, shifts across cross-sectional units, and correlations over time and space. Such multiple imputations are performed with a bootstrap based EMB algorithm, by inferring the missing values based on yearly trends in each region, treating the time-series-cross-sectional nature of the datum.

\subsubsection{GDP per Capita}
\label{var gdp}
We select GDP per capita as it is one of the most universally adopted and straightforward indicators of the overall economic wealth of a region. In order to compare values for different nations, we correct raw GDP per capita with the Purchasing Power Parity (PPP) proposed by The World Bank \cite{world_bank}.
We collect available tables for regional GDP per capita from different sources which are reported in Appendix \ref{appendix a: gdp}.

Due to the absence of a full table containing regional data, we obtain our own with two strategies. Data from different origins are not comparable due to differences in currency and Purchasing Power corrections, for this reason the collected values are corrected such that the national year-wise GDP per capita is coherent with the one reported in the World Bank's national tables. This operation consists in the computation, for every country and for every year, of a proportionality coefficient between the national data reported by The World Bank and the one reported in the different tables, and a subsequent application of said proportionality coefficients across all regions of said country. Using the same PPP correction uniformly across a whole nation, albeit being an approximation, is a necessary one. Moreover, it is a standard practice in such scenarios in the econometric practice. \cite{reg_eco10}.
Since some data are still missing after this procedure, we impute them through \textit{Amelia}.
A deeper explanation on pre-processing and imputation procedure can be found in Appendix \ref{appendix a: gdp}.


\subsubsection{Education Index}
\label{var edu}
As an indicator of the overall schooling provision of a region we select the Education Index. This indicator is a component of the Human Development Index and it is defined as the geometric average of mean years of schooling and expected years of schooling \cite{Saisana_edu_index_2014}. Sources for the tables we use are reported in Appendix \ref{appendix a: edu index}.

The basis for the data is taken from Global Data Lab, however regional NUTS2 data is not available for all countries and imputation is exploited. Such imputation is performed using data from Eurostat, which reports the percentage of tertiary education attainment in the 25-64 age groups, modeling the correlation between this value and the Education Index. This procedure is rendered necessary since all imputation methods exploit the correlation structure of the dataset, whether it is correlated regions or auto-correlated years, but in a situation where the datum is missing as a whole for entire countries we need to exploit other types of correlation.
Some data is still missing at this point, notably no datum was available for the Norwegian region \textit{NO0B}, Jan Mayen and Svalbard, for which the national average was applied. Other cases are again imputed through \textit{Amelia}. 
More details on the extraction, pre processing and imputation procedure are reported in Appendix \ref{appendix a: edu index}.

\subsubsection{University Score}
\label{var uni}
We construct an indicator of the overall prestige of a region's academic institutions, by considering the university ranking proposed each year by QS \cite{aboutQS}, table sources are reported Appendix \ref{Appendix  A - uniscore}.

The way we compute it is by attributing to each university in the top 500 positions in the global ranking a weight determined by its position in the ranking. The weight attribution mechanism is detailed in Table \ref{tab:bands_US_weights}, defining weights as the inverse of half the lower bound of the respective band.

\begin{table}[t]
\begin{center}
\caption{\label{tab:bands_US_weights}Weight attributed to universities in QS ranking.}
\begin{tabular}{@{}ll@{}}
\toprule
Position in ranking& Weight\\
\midrule
1-10&1/5\\
11-20&1/10\\
21-50&1/25\\
51-100&1/50\\
101-250&1/125\\
251-500&1/250\\
\bottomrule
\end{tabular}
\end{center}
\end{table}

Once each university has been given a weight on each year's ranking we compute the University Score of a region in a given year as the cubic root of the sum of weights of all the universities in that year's top 500 ranking which are in said region. 
\[
University\_Score_{year}(region)= \sqrt[3]{ \sum_{uni = 1}^{500} \left(weight_{year}(uni)*\mathbbm{1}(uni \in region)\right) }
\]
The cubic root had been introduced in order to generate a more uniformly distributed indicator across regions.
For universities located in multiple regions, we consider them as being only in the region reported on the official QS website \cite{aboutQS}.

Although the construction of this indicator is arbitrary, it accounts for the number of prestigious institutions while heavily preferring top ranked ones, therefore we believe it to be a representative indicator of the overall academic prestige of a region.

\subsubsection{TED}
\label{var ted}
We construct an indicator of the overall amount of investments present in each region on top of Tenders Electronic Daily (TED) public procurement notices. Details on the table used are reported in Appendix \ref{appendix a: ted}.

From the notices table, we filter for the procurement whose value has actually been awarded, and we consider the correspondent NUTS2 region, inferred from the postal code if the datum was missing, the year of reference and the value in euro.

We build our TED indicator as the natural logarithm of the sum of all awarded capital in each given region, in each year.

Notice that data has to be inserted compulsorily into the register only if the value of the procurement is above the procurement threshold, which depends on both type of contract and contracting authority, even though publishing below threshold tenders in TED registers is considered a good practice. Additionally, since data in the register may have been incorrectly inserted or even be missing, not all procurement notices are successfully mapped into NUTS2 regions. For these reasons, when faced with regions where, for a given year, no procurement notices are attributed, we decide to treat this as a missing datum to be imputed, instead of a complete absence of capital investments in the region. More information on the pre-processing applied and imputation performed are reported in Appendix \ref{appendix a: ted}.

Even though the procurement in TED registers do not specifically target research, we use the order of magnitude of capital emittance as an indicator of the overall amount of public investments available in each given region.


\section{Preliminary Analysis}
\label{section: Preliminary Analysis}
We present a preliminary exploratory analysis conducted on the dataset extracted in Section \ref{subsection: ORCID data} and a characterization of its dependency over time-space dominion.

Section \ref{subsection: Data Exploration} briefly illustrates the composition of our dataset, together with summary statistics, while Section \ref{subsection: Temporal Characterization} and Section \ref{subsection: Spatial Characterization} respectively explore the temporal and spatial dependencies of our problem.

Results are coherent with national ones, briefly described in Appendix \ref{Appendix C national preliminary analyisis}.

\subsection{Data Exploration}
\label{subsection: Data Exploration}

ORCID public data contains informations on \numprint{3722294} individual ORCID users, of which \numprint{861752} have either started or ended at least one affiliation in a region of interest in the 2009-2020 time horizon and of said population approximately 40\% has a PhD. 

The information published in the ORCID registers is manually reported and its publication is voluntary, as a consequence approximately only 60\% of the researchers we track for our analysis have reported their education affiliation as well. We decide to include education experiences (whether they are Bachelors, Masters or PhDs) in our analysis, tracking the migrations that occurred in said periods too. We indeed believe that such migrations consist of an integral part of the phenomenon, as the student mobility which we are tracking refers to a specific and relevant subset, being composed by people that do take part in research and innovation later on in their career.

\subsection{Temporal Characterization}
\label{subsection: Temporal Characterization}

Our analysis tracks migrations over the span of 12 years, this implies not only possible temporal heterogeneities in the phenomenon, but a time-based correlation structure in our observations as well.

\begin{figure}
\centering
\caption{\label{figure:tot flow on years} Total regional mobility by year.}
\makebox{\includegraphics[scale=0.35]{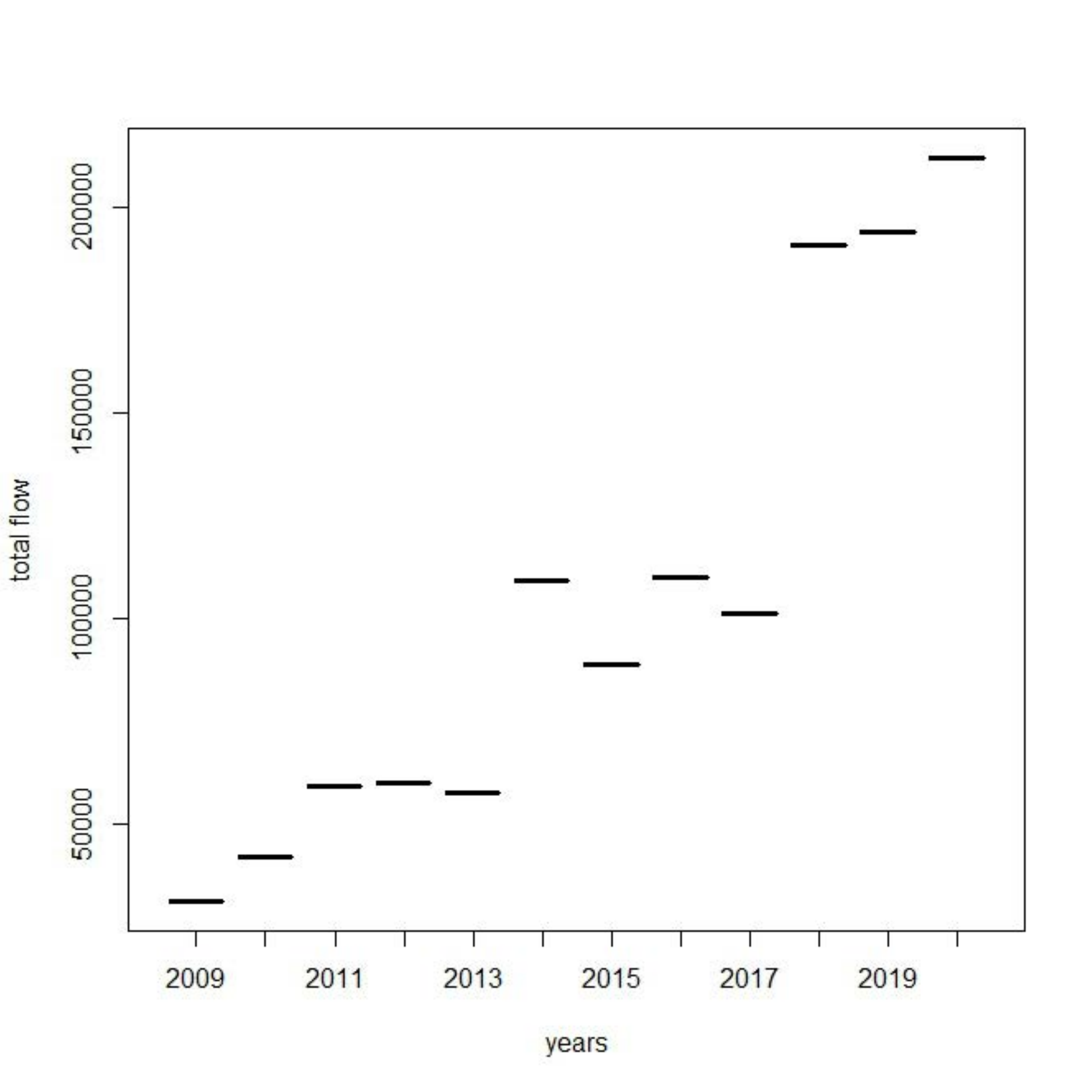}}
\end{figure}

We start by addressing the heterogeneity induced by time, which can be modeled into two components that should be considered differently: an inflation factor on mobility across different years and an actual shift in the migration patterns.\\
We notice strong magnitude differences year by year of the total regional mobility, defined as the total number of people coming in a region plus leaving said region, as shown in Figure \ref{figure:tot flow on years}.
While different migration patterns can be attributed to changes in the nature of the phenomenon, we are careful about interpreting magnitude changes that occur uniformly across the network, as they could just be caused by different adoption rates of ORCID.
Such a consideration will always be present during our analysis, in which we try to study the phenomenon up to a scale coefficient across different years.

\subsubsection{Principal Components Analysis}

Principal components analysis (PCA) is a standard practice in all of machine learning to study the variability of a piece of data across its components.

We analyse the total number of researchers leaving a region plus the total number of researchers entering said region (total mobility), over all 12 years.\\
We perform a PCA considering as data each regions, over the 12 dimensions induced by time, both for the total flows and for the yearly standardized flows, as normalizing data is a common practice when exploiting PCA. We obtain poorly interpretable principal components and low variability explained by each one in both cases. A summary description of the results is reported in Table \ref{tab:PCA}. Very similar results come from the same analysis performed on the total number of researchers leaving a region and on the total number of researchers entering a region, year by year. Details can be found in Appendix \ref{Appendix D pca in out}.\\
Such findings indicate a heterogeneous nature of the phenomenon across time. Whether such heterogeneity is explainable in terms of fluctuations of regional characteristics, it is purely quantitative and imputable to the increasing popularity of ORCID or it dictates a shift in the phenomenon nature will be a topic explored over the course of the whole analysis. The results of Section \ref{section: Mobility Analysis} and Section \ref{section: Models} in particular are investigated in light of this finding.

%

\begin{table}[t]
\begin{center}
\caption{\label{tab:PCA}PCA summary for the first 3 principal components  (PCs).}
\begin{tabular}{@{}c|*{3}{r}|*{3}{r}@{}}
\toprule
&\multicolumn{3}{c|}{ Total Flows}& \multicolumn{3}{c}{ Normalized Flows} \\
& \textbf{1st PC} & \textbf{2nd PC} & \textbf{3rd PC}& \textbf{1st PC} & \textbf{2nd PC} & \textbf{3rd PC}\\\hline\hline
 Proportion of variance explained & 0.338 & 0.265 & 0.181 & 0.198  & 0.117 & 0.099 \\\hline
 Year &\multicolumn{6}{c}{ Component's loading}\\\hline

2009 &      &       &       &      0.268&  0.149&  0.354\\   
2010 &      &       &       &      0.237& -0.185& -0.378\\   
2011 &      &       &       &      0.347&  0.461& -0.190\\   
2012 &      &       &       &      0.201& -0.311&       \\   
2013 &      &       &       &      0.330&       &  0.492\\   
2014 &      &  0.112&       &      0.147&       & -0.361\\   
2015 &      &       &  0.160&      0.408& -0.343&       \\
2016 &      &       &       &      0.283&  0.143&  0.281\\   
2017 &      &       &       &      0.337& -0.126&  0.277\\   
2018 & 0.943& -0.316&       &      0.222&  0.369& -0.152\\   
2019 & 0.292&  0.927& -0.174&      0.239&  0.365& -0.307\\   
2020 &      &  0.143&  0.95 &     0.339 &-0.447 &-0.194\\
\bottomrule
\end{tabular}
\end{center}
\end{table}

\subsection{Spatial Characterization}
\label{subsection: Spatial Characterization}

We now explore the spatial dimension of our problem, trying to identify groups of regions based on their connectivity.

\subsubsection{Community Detection}

We now model the phenomenon via a fully connected, bidirectional weighted network, where nodes are regions and arc weights are the total amount of researchers, summing across the whole time horizon, leaving one region for another one. We refer to this as cumulative network during this discussion. Due to the possible time heterogeneity highlighted in Section \ref{subsection: Temporal Characterization}, the following analysis is validated by assessing the consistency of results across all 12 different networks, where arc weights are computed only on top of the researchers traveling in the specific year.

There is plenty of community detection methods which try to identify grouping structure amongst nodes, based on different definitions, we present two: edge betweenness community detection \cite{newman_finding_2004} and infomap community detection \cite{rosvall_maps_2008}. Both methods are suitable for bidirectional weighted networks like the one induced by the problem, but their method for defining communities lead to different conclusions.

Edge betweenness community detection \cite{newman_finding_2004} identifies communities as densely connected sub-graphs which are sparsely connected between them. Such a method applied on our cumulative network aggregates all nodes into a single community and does so on all year-wise network too. This leads us to the idea that the phenomenon by itself is quite uniformly connected and there are no evident choke points or isolated communities in the migration pattern.

\begin{figure}
\centering
\caption{\label{figure:random walk division} Partition induced by infomap clustering.}
\makebox{\includegraphics[scale=0.4]{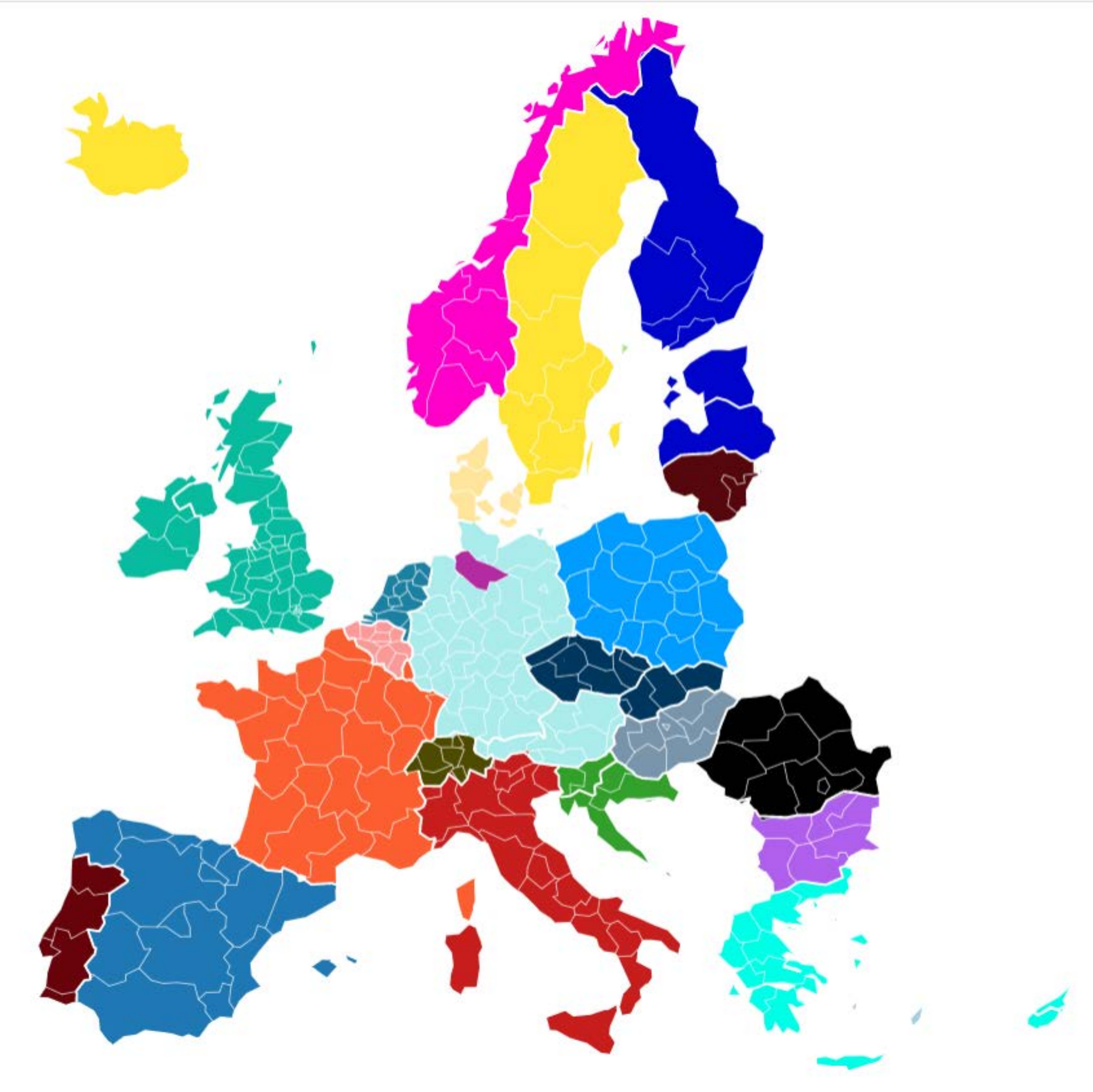}}
\end{figure}

A different approach is the one offered by infomap community detection \cite{rosvall_maps_2008}, which defines communities based on the expected frequency of visit of nodes of a random walker on the graph. This method, as opposed to the previous one, is able to highlight communities in a quite uniformly connected network like the one we are working on. Indeed the results paint a picture where regions are grouped based on national borders, sometimes aggregating together exceptionally close countries. National borders are respected in all year-wise networks as well. A visualization of the detected community partition is reported in Figure \ref{figure:random walk division}.

Albeit the nature of the researchers' migrations phenomenon is not inherently constrained by national borders, and the results of edge betweenness community detection confirm it, there still is an important component of regional migrations inside countries, as highlighted by infomap community detection.
For this reason we can expect a hierarchical dependency structure to exist between regions belonging to the same country. This consideration will be taken into account in the models proposed in Section \ref{section: Models}.

\section{ANOVA with NPCs}
\label{section: ANOVA NPC}

In Section \ref{subsection: Network Nodes Analysis} we will present partitions on our regions, with the goal of testing if said partitions are reflected in the regional characteristics presented in Section \ref{subsection:coovariates}.
We adjust a robust, non parametric version of the binary ANOVA test to provide detailed results on the differences between two groups of observations.

Non parametric ANOVA tests indeed assume under the null hypothesis H0 an identical distribution between the two groups. In our analysis, however, we are not only interested in a generic discrepancy in distribution, instead we want to assess what causes the violation of said hypothesis, meaning different centers of the populations' distributions or differences in the dispersion metrics. Our solution is built on top of multi-aspect testing theory \cite{pini_multi-aspect_2019}, with a p-value correction offered by non parametric combinations theory \cite{caughey_nonparametric_2017}, and detailed over the course of this Section. 

\subsection{Theoretical Foundations}
\label{subsection: NPC theo}
The idea of multi-aspect testing consists in decomposing the original test into partial ones, each focusing on a specific violation of the null hypothesis. Ultimately such partial tests, in order to still control type I error, are in need of a p-value correction. Such p-value correction is performed though non parametric combinations, which consist of correcting the p-value of multiple tests by means of joint ones.

The procedure can hence be summarised in the following steps:
\begin{enumerate}
    \item From the null hypothesis of equality in distribution between the two groups, define two tests, one whose alternative hypothesis is difference in centers, the other whose alternative hypothesis is heteroscedasticity. Obtain p-values of the tests: $\text{p-value}_{location}$ and $\text{p-value}_{scale}$.
    \item Define a third test, whose alternative hypothesis is the union of the two, hence difference in location or scale. Obtain p-value of the test: $\text{p-value}_{joint}$.
    \item Correct the p-values according to the formula:
    \[
    \text{corrected p-value}_i = max(\text{p-value}_i,\  \text{p-value}_{joint}) \quad i \in \{location, scale\}
    \]
\end{enumerate}

Note that for all three tests the null hypothesis is the same, meaning equality in distribution. Non parametric theory easily allows to define them, as it suffices to appropriately define test statistics which are particularly susceptible to the violation expressed by the alternative hypothesis.

\subsection{Test Definition}

We now show how we put in practice the framework outlined in Section \ref{subsection: NPC theo}. We start by introducing the location test, detailed in Test (\ref{test location}).
\begin{eqnarray}
\label{test location}
&&H0: \mathcal{L}(g_1) = \mathcal{L}(g_2) \qquad H1: \mathbbm{E}[\mathcal{L}(g_1)] \neq \mathbbm{E}[\mathcal{L}(g_2)]\nonumber\\
&&Test \ statistic = \left(Median(g_1) - Median(g_2)\right)^2\\
&&Permutation \ scheme: permutation \ of \ observations \nonumber
\end{eqnarray}
With $\mathcal{L}(g)$ being the law of group $g$ and $\mathbbm{E}[\mathcal{L}(g)]$ its center.
The median is used as location metric due to its robustness.
Directly permuting observations is a likelihood-invariant permutation scheme under H0.

The scale test is presented as Test (\ref{test scale}).
\begin{eqnarray}
\label{test scale}
&&H0: \mathcal{L}(g_1) = \mathcal{L}(g_2) \qquad H1: Var[\mathcal{L}(g_1)] \neq Var[\mathcal{L}(g_2)]\nonumber\\
&&Test \ statistic = \left(MAD(g_1) - MAD(g_2)\right)^2\\
&&Permutation \ scheme: permutation \ of \ observations \nonumber
\end{eqnarray}

Being $Var[\mathcal{L}(g)]$ the dispersion parameter of group $g$'s distribution and $MAD(g)$ the median absolute deviation of group $g$, used as a robust metric for the variability within said group.

In order to provide the p-value correction, we define the joint test as Test (\ref{test joint}).
\begin{eqnarray}
\label{test joint}
&&H0: \mathcal{L}(g_1) = \mathcal{L}(g_2) \qquad H1: (\mathbbm{E}[\mathcal{L}(g_1)] \neq \mathbbm{E}[\mathcal{L}(g_2)]) \cup (Var[\mathcal{L}(g_1)] \neq Var[\mathcal{L}(g_2)]) \nonumber\\
&&Test \ statistic = max\{max(U_1,U_2), max(V_1,V_2)\}\\
&&Permutation \ scheme: permutation \ of \ observations \nonumber
\end{eqnarray}

Being $U_1$, $U_2$ the two Mann-Whitney U-statistics \cite{mann_test_1947} computed via sum of ranks of the direct observations of the two groups, and $V_1$, $V_2$ the two Mann-Whitney U-statistics computed via sum of ranks over the square deviation of the observations with respect to their group median. Such a choice for a test statistic is crucial for the definition of Test (\ref{test joint}), indeed we want the test statistic to be equally susceptible to violations of the null hypothesis for both location and scale imbalances. A definition based on rankings removes the dimensionality of location and scale metrics, allowing for a balanced comparison.

After having performed Tests (\ref{test location}), (\ref{test scale}) and (\ref{test joint}) and having applied the correction defined in Section \ref{subsection: NPC theo}, $\text{p-value}_{location}^{corrected}$ and $\text{p-value}_{scale}^{corrected}$ offer a robust result for testing whether there is a significant difference in the centers and scale parameters of the distributions of two univariate populations.

\section{Mobility Characterization}
\label{section: Mobility Analysis}

This Section tries to address the nature of researchers migration and answer the question over the existence of a brain drain phenomenon, restricted to our time-space horizon. Section \ref{subsection: Correlation Study} explores the correlation between flows of researchers entering or leaving a region, while \ref{subsection: Network Nodes Analysis} explores mobility under a network perspective.

Results are coherent with national ones, reported in Appendix \ref{Appendix C national mobility characterization}.

\subsection{Correlation Study} 
\label{subsection: Correlation Study}

For each region, in each year, we compute the total amount of researchers entering said region or leaving it. We refer to these quantities as "in flow" and "out flow". 
Section \ref{subsub: raw correlation} shows an exceptional correlation between the two quantities, albeit computed in an aggregated fashion. Section \ref{subsub: Spearman} provides a more robust support to the claim by working in a functional framework.

\subsubsection{Raw Correlation}
\label{subsub: raw correlation}
We now consider the total in flow and total out flow summed over the whole time horizon for each region. The raw correlation between total in flow and total out flow is 98.97\%. We build a simple ordinary least squares linear model and a least trimmed squared robust linear model to account for possible outlying points, as some regions present an extremely high flow with respect to others. We obtain two almost identical regression lines, with the one obtained by ordinary least squares being:
\[
total\_out\_flow = 8.803155 + 0.99*total\_in\_flow 
\]
with such model having a R-squared of 0.979.

Such result shows a strong correspondence between the total incoming flow in a region and that region's outgoing flow, but does not take into account the repeated measure structure of our observations. Such effects are considered in the next Section.

\subsubsection{Spearman Correlation}
\label{subsub: Spearman}

In order to account for the time-dependent nature of our observations, we redefine the datum to be a region, considering the 12 repeated measures of its total in flow and total out flow as being twelve measurements from the same functional datum. Raw Spearman correlation index \cite{valencia_garcia_spearman_2013} between the two functional data is 0.98. 

After performing a smoothing over the discrete year-wise measurements, the two-dimensional functional datum is composed of the total in flow and total out flow trends over time, for each individual NUTS2 region. Spearman correlation test \cite{valencia_garcia_spearman_2013}, illustrated in Test \ref{test spearman}, allows to assess the correlation between the two functional dimensions of the datum.

\begin{eqnarray}
\label{test spearman}
&&H0: \rho_s(\textbf{in\_flows}, \textbf{out\_flows}) = 0 \qquad H1: \rho_s(\textbf{in\_flows}, \textbf{out\_flows}) \neq 0 \nonumber\\
&&Test \ statistic = |\rho_s(\textbf{in\_flows}, \textbf{out\_flows})|\\
&&Permutation \ scheme: permutation \ of \ \textbf{out\_flows} \ observations \nonumber
\end{eqnarray}

With \textbf{in\_flows} and \textbf{out\_flows} being the vectors containing region-wise functional data on flow trends and $\rho_s$ being Spearman's correlation index for functional data.
With a p-value of 0 we confirm the strong dependence between the two variables.

These are the first results supporting the introduction of a different characterization for researcher's migration with respect to brain drain, which we will generally refer to as "brain mobility". We indeed notice that researchers are not leaving \emph{en masse} some specific regions to migrate into others, but the regions in our domain seem to be instead characterizable in terms of a high or low exchange of talents.

\subsection{Network Nodes Analysis}
\label{subsection: Network Nodes Analysis}
In this Section we try to explore the idea of brain mobility and we aim at finding the most relevant nodes on the cumulative network and on the year-wise networks previously defined in Section \ref{subsection: Spatial Characterization}. In Section \ref{subsub: Hubs and Authorities Analysis}, we try to identify hubs and authorities of the networks \cite{Kleinberg1999AuthoritativeSI}: hubs are provider regions, they export many researchers in direction of the most attractive countries; authorities are attractors, they draw in researchers from the most providing regions. Afterwards, Section \ref{subsub: S-coreness Analysis} exploits coreness analysis \cite{eidsaa_s-core_2013} to identify all groups of regions in which each region has at least a fixed value of weighted degree, in order to find the most connected ones.

\subsubsection{Hubs and Authorities Analysis}
\label{subsub: Hubs and Authorities Analysis}
We work with a weighted bidirectional fully connected network, with nodes being the regions of our spatial domain, and our goal is to identify possible regions acting as strong providers and receivers of researchers, namely hubs and authorities. We do so on the cumulative network, where arc weights are the total number of researchers migrating from a region to another across the whole time horizon, and confirm the results on the 12 year-wise networks, where we only consider the migrations that happened in each specific year separately.

HITS (hyperlink-induced topic search) algorithm was developed in \cite{Kleinberg1999AuthoritativeSI}, with the objective of ranking Web pages, defining authorities as the most cited pages and hubs as the most citing ones, but it can be easily generalized to different domains like the one we are working on \cite{urbinati_measuring_2021}, interpreting hubs and authorities as providers and attractors of talents, respectively.

\begin{figure}
\centering
\caption{\label{figure:aut vs hubs}Authorities and hubs scores correspondence.}
\makebox{\includegraphics[scale=0.45]{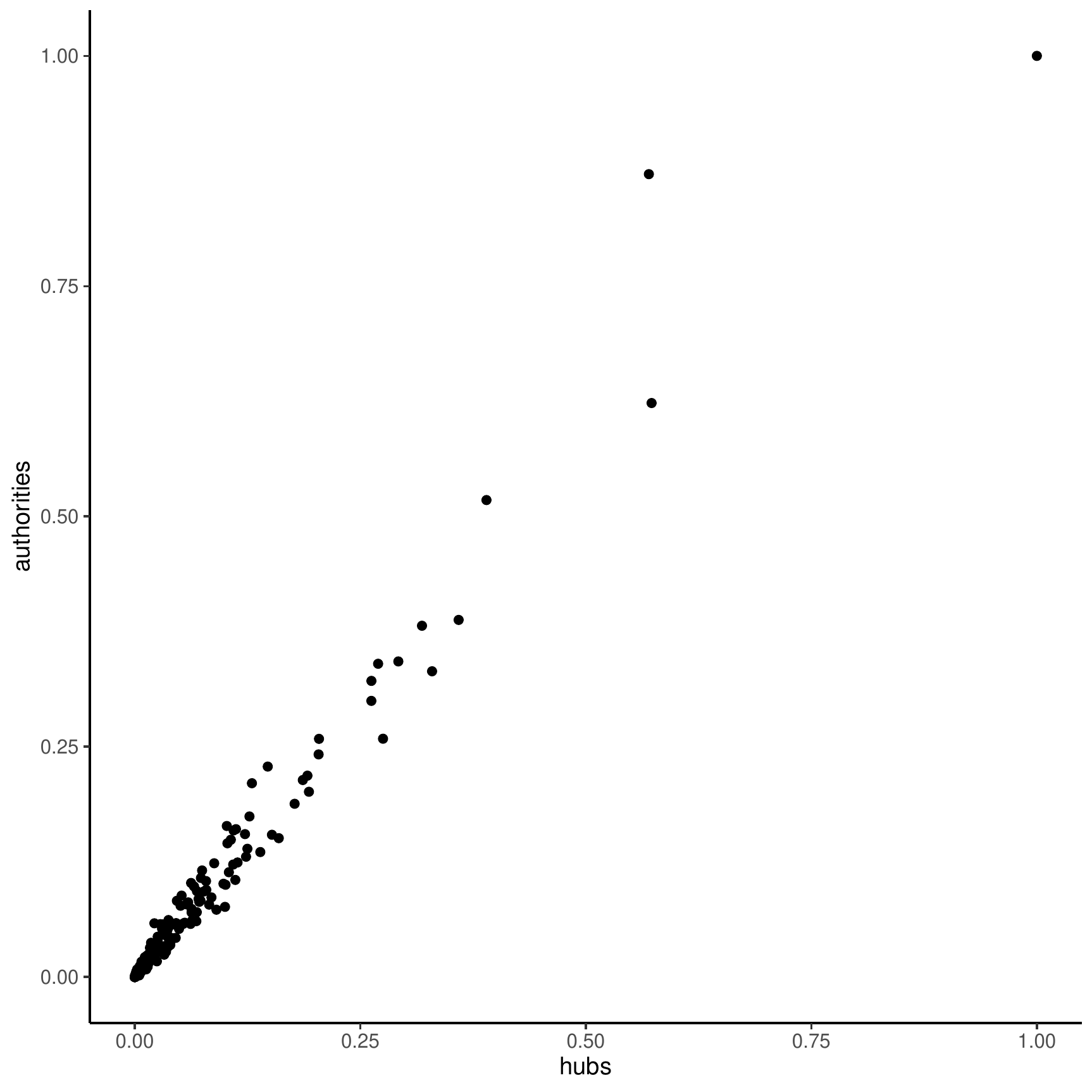}}
\end{figure}

We apply the algorithm on the cumulative network and obtain a positive correlation of 98\% between the hubs and authorities scores. Such result suggests that there are no net providers or attractors of talent in the network, but the regions who attract more talents are also equally important exporters, further supporting the results of Section \ref{subsection: Correlation Study}.\\
The high correspondence between the two scores is visible in Figure \ref{figure:aut vs hubs}, while the correlation results for year-wise networks are reported in Table \ref{tab: HubsAut corr}. It is possible to notice how, especially in the latest years of the analysis, there is an exceptional positive correlation between such scores, while in year 2010 such correlation is weaker. We do not have the means to evaluate whether the outlying behaviour of 2010 is due to the high sparsity of data in the first years of analysis, particular events that took place in said year or a mix of both, we decide however to treat it as an irregular observation and keep working on the cumulative network, while still being aware of the time heterogeneity.

\begin{table}[t]
\begin{center}
\caption{\label{tab: HubsAut corr}Correlation between hubs and authorities scores of year wise networks' nodes.}
\begin{tabular}{@{}cc@{}}
\toprule
 Year &  Correlation\\
\midrule
2009& 77\% \\
2010& 42\% \\
2011& 75\% \\
2012& 93\% \\
2013& 95\% \\
2014& 96\% \\
2015& 94\% \\
2016& 96\% \\
2017& 97\% \\
2018& 97\% \\
2019& 98\% \\
2020& 97\% \\
\bottomrule
\end{tabular}
\end{center}
\end{table}

\begin{figure}
\centering
\caption{\label{figure:hubs aut hist}Histograms of hubs (A) and authorities (B) scores on cumulative network.}
\makebox{\includegraphics[scale=0.45]{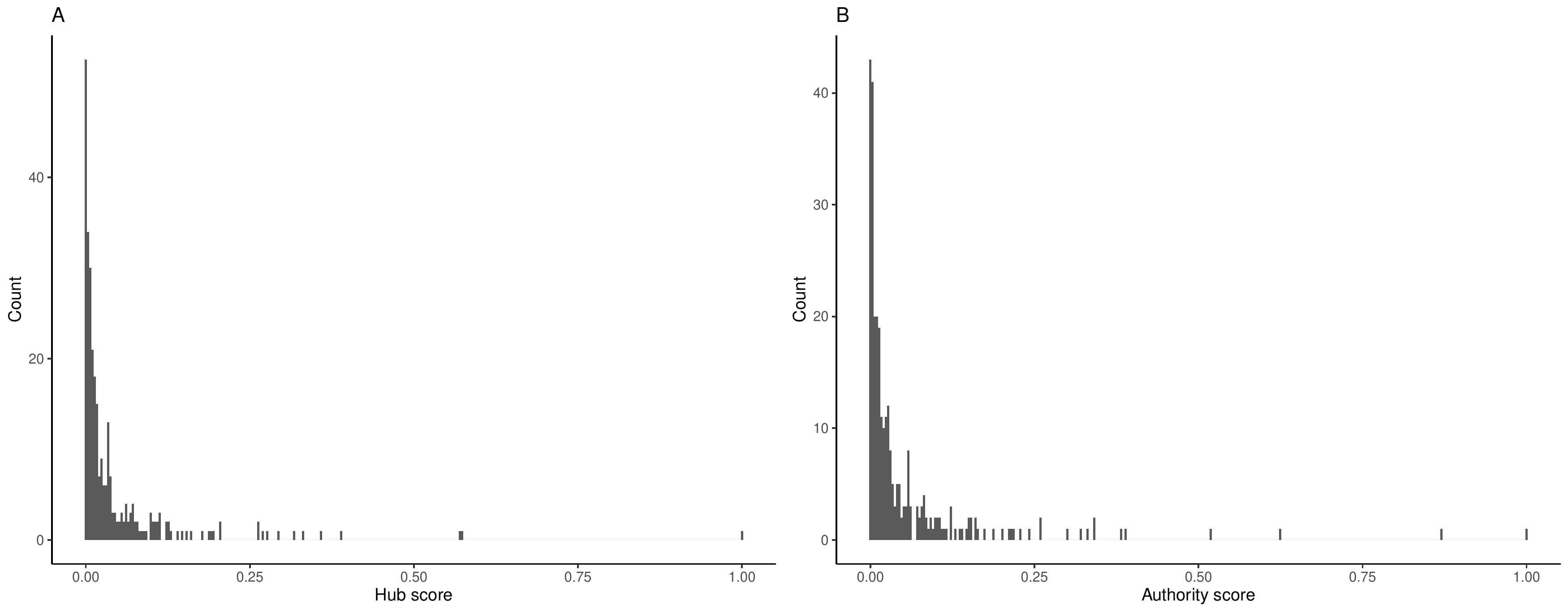}}
\end{figure}

Given the scores, we aim at meaningfully partitioning regions based on their roles and see how regional characteristics may differ between the groups.
Being the scores generally continuous and very concentrated near zero, as shown in Figure \ref{figure:hubs aut hist}, with the exception of a few observation with extremely high scores, we try two different cutting points, namely at the boundaries of said extreme observations and at the 90\% quantile on scores distributions. On the induced partitions we then verify, with the NPC-based ANOVA test presented in Section \ref{section: ANOVA NPC}, that the regions' characterizing variables indeed behave differently between high scores regions and low scores regions.

\begin{figure}
\centering
\caption{\label{figure:hubs aut 0.25} Regions with exceptionally high hubs and authority scores (darker) compared to other regions (lighter).}
\makebox{\includegraphics[scale=0.4]{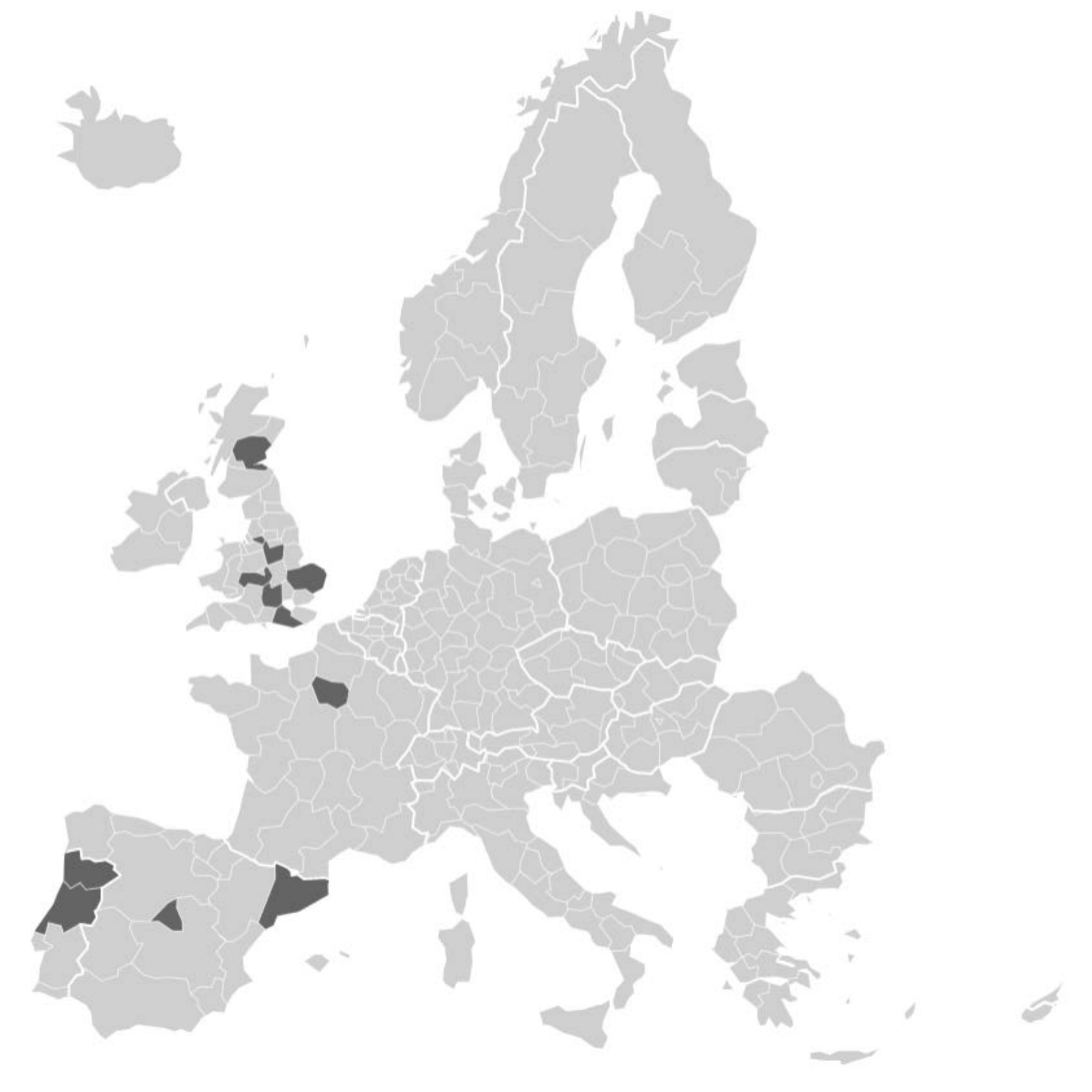}}
\end{figure}

The extreme observations are determined as those which have hubs or authority score higher than 0.25, with the English region of Derbyshire and Nottinghamshire, \textit{UKF1}, being the only one which is extreme in authority score and not in hub score. A representation of the induced partition is reported in Figure \ref{figure:hubs aut 0.25}, the results of the NPC-based ANOVA tests on said partition are reported in Table \ref{tab:ANOVA HA outl}.

\begin{table}[t]
\begin{center}
\caption{\label{tab:ANOVA HA outl}ANOVA test results for partition induced by exceptionally high hubs and authorities scores.}
\begin{tabular}{@{}ccc@{}}
\toprule
 Variable& Location p-value& Scale p-value\\
\midrule
GDP per capita & 0.210 & 0.679\\
Education Index & 0.130 & 0.586\\
TED & 0.000 & 0.926\\
University Score & 0.001 & 0.001\\
\bottomrule
\end{tabular}
\end{center}
\end{table}

Even for such a fine partition, being one group composed by only 13 observations, it is possible to gather evidence for a meaningful difference of regional characteristics, especially University Score and TED.

\begin{figure}
\centering
\caption{\label{figure:hubs aut hist 0.9} Regions with hubs and authority scores above (darker) and below (lighter) the 90\% quantile of the respective distribution.}
\makebox{\includegraphics[scale=0.4]{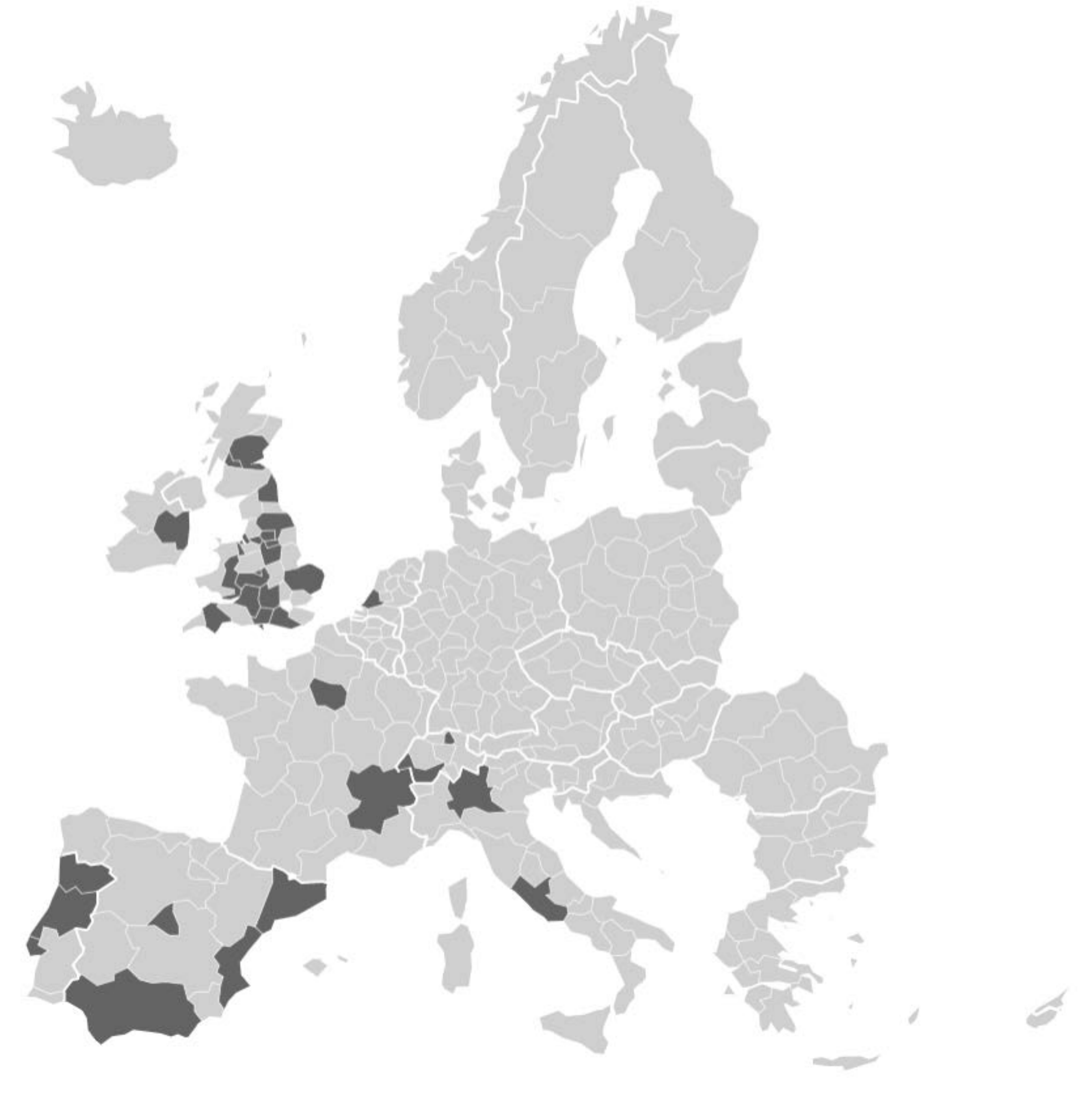}}
\end{figure}

We also test a quantile induced partition, meaning that we define as high mobility the union of regions with top 10\% hub or 10\% authority scores, strong of the correlation between the two. A visualization of said partition is reported in Figure \ref{figure:hubs aut hist 0.9}, the results of the NPC-based ANOVA test on said partition are reported in Table \ref{tab:ANOVA HA 0.9}.

\begin{table}[t]
\begin{center}
\caption{\label{tab:ANOVA HA 0.9}ANOVA test results for 90\% quantile hubs and authorities scores induced partition.}
\begin{tabular}{@{}rrr@{}}
\toprule
 Variable& Location p-value& Scale p-value\\
\midrule
GDP per capita & 0.080 & 0.315\\
Education Index & 0.030 & 0.190\\
TED & 0.000 & 0.854\\
University Score & 0.000 & 0.000\\
\bottomrule
\end{tabular}
\end{center}
\end{table}

Similarly to the previous partition, the difference in mobility is highly reflected on University Score and TED. 

We check that the 90\% quantile partition is uniform in the year-wise networks as well, by dividing the nodes at the 90\% quantile of hubs and authorities scores for each year separately and computing the accuracy of groups belonging with respect to the cumulative network scores division. These can be interpreted as congruence scores between the grouping structures found for the 12 networks with respect to the cumulative network one: we obtain a correspondence between 91\% and 98\% for hubs, between 92\% and 98\% for authorities scores. This indicates that for every year this partition remains consistent and that regions with high hubs and authorities scores usually stay the same in time. We also check whether this partition remains significant for year-wise regional characteristics, results of the respective tests are reported in Appendix \ref{appendix e: hubs aut yearly anova} and show that especially the behavior of University Score and TED remains consistent in time.

\subsubsection{$S$-Coreness Analysis}
\label{subsub: S-coreness Analysis}

We now work with the same networks studied in Section \ref{subsub: Hubs and Authorities Analysis} and our goal is to define a possible core of the network, after discovering a strong correspondence between hubs and authorities regions.

We use the following notation: we have \textit{N} nodes (regions) and \textit{M} links, the adjacency matrix is defined as $A=[a_{ij}]$, where $a_{ij} = 1$ when nodes $i$ and $j$ are connected, zero otherwise, finally the link-weight matrix, indicating strength of a connection $w_{ij}$, is $W=[w_{ij}]$.\\
In the $s$-core decomposition of a network \cite{eidsaa_s-core_2013}, the \textit{s}-core consists of all nodes $i$ with node strengths $s_{i} > s$, where s is a threshold value and $s_{i}$ is the weighted strength of node $i$, either considering incoming edges, outgoing edges or the sum of the two. The threshold value of the $s_{n}$-core is defined as $s_{n-1} = \min_{i} s_{i}$, where $i$ is only among the nodes in the $s_{n-1}$-core network. The $s_{n}$-core is thus identified by the iterative removal of all nodes with strengths $s_{i} \leq s_{n-1}$.

We present the analysis for the cumulative network over years with the strength of a node $i$ being $s_{i}=\sum_j a_{ij}w_{ij} + \sum_j a_{ji}w_{ji} $, namely the total sum of researchers entering a region plus the total sum leaving said region, with strengths being iteratively computed on the sub-network identified by the previous core at each iteration.\\
We validate our results by using different definitions for the strength of nodes, namely $s_{i}=\sum_j a_{ij}w_{ij}$, representing the total outgoing flow from a region, and $s_{i}= \sum_j a_{ji}w_{ji} $, describing the total incoming flow in a region, again iteratively computed on the sub-networks identified by the previous core. These other analyses are reported in Appendix \ref{Appendix D s-coreness in out} and are coherent with the first case, further confirming the correlation studied in Section \ref{subsection: Correlation Study}.

\begin{figure}
\begin{minipage}[c]{0.45\linewidth}
\caption{\label{figure: s-core distr} $s$-cores thresholds distribution.}
\makebox{\includegraphics[scale=0.32]{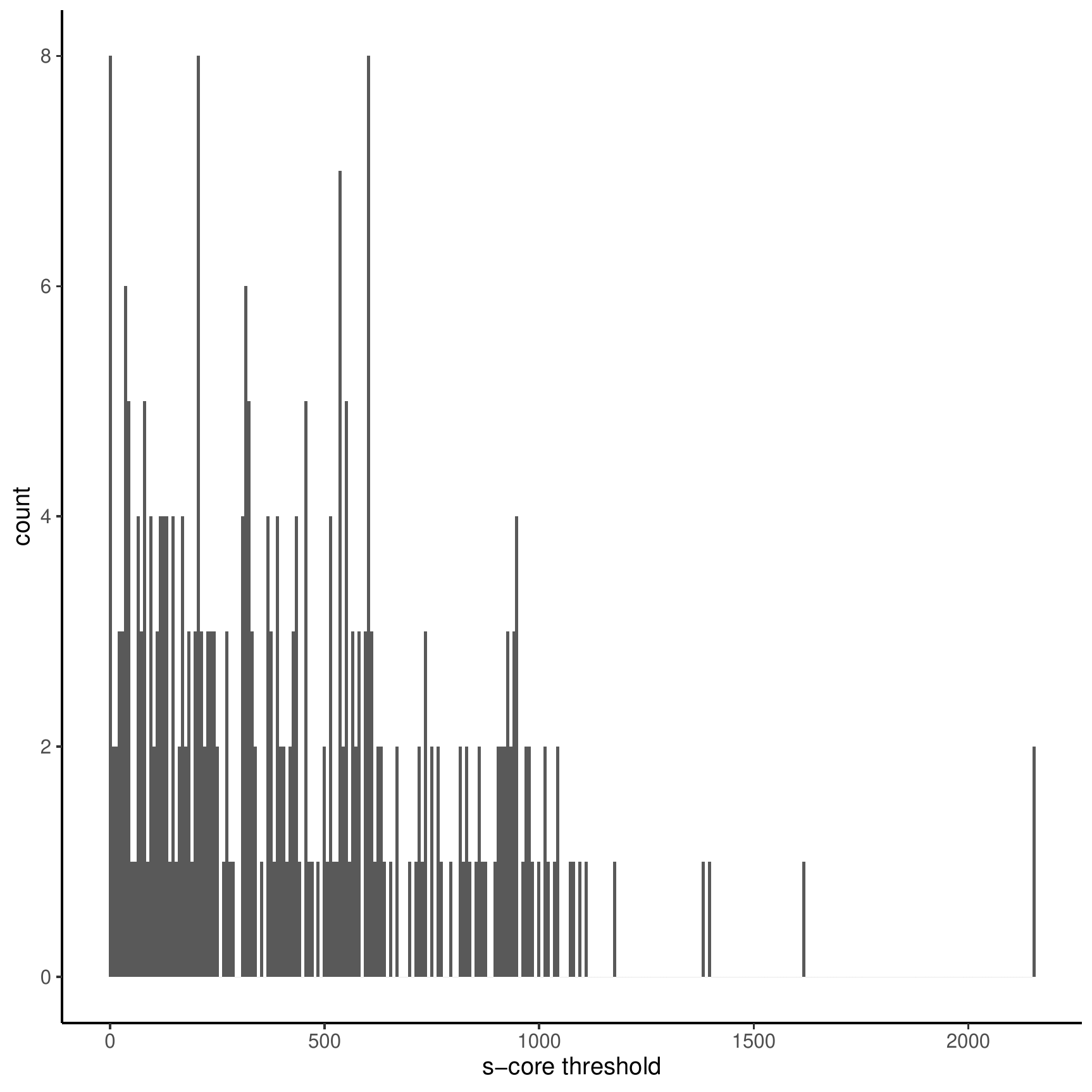}}
\end{minipage}
\hfill
\begin{minipage}[c]{0.45\linewidth}
\caption{\label{figure: s-core shell} $s$-cores shells numerosity.}
\makebox{\includegraphics[scale=0.32]{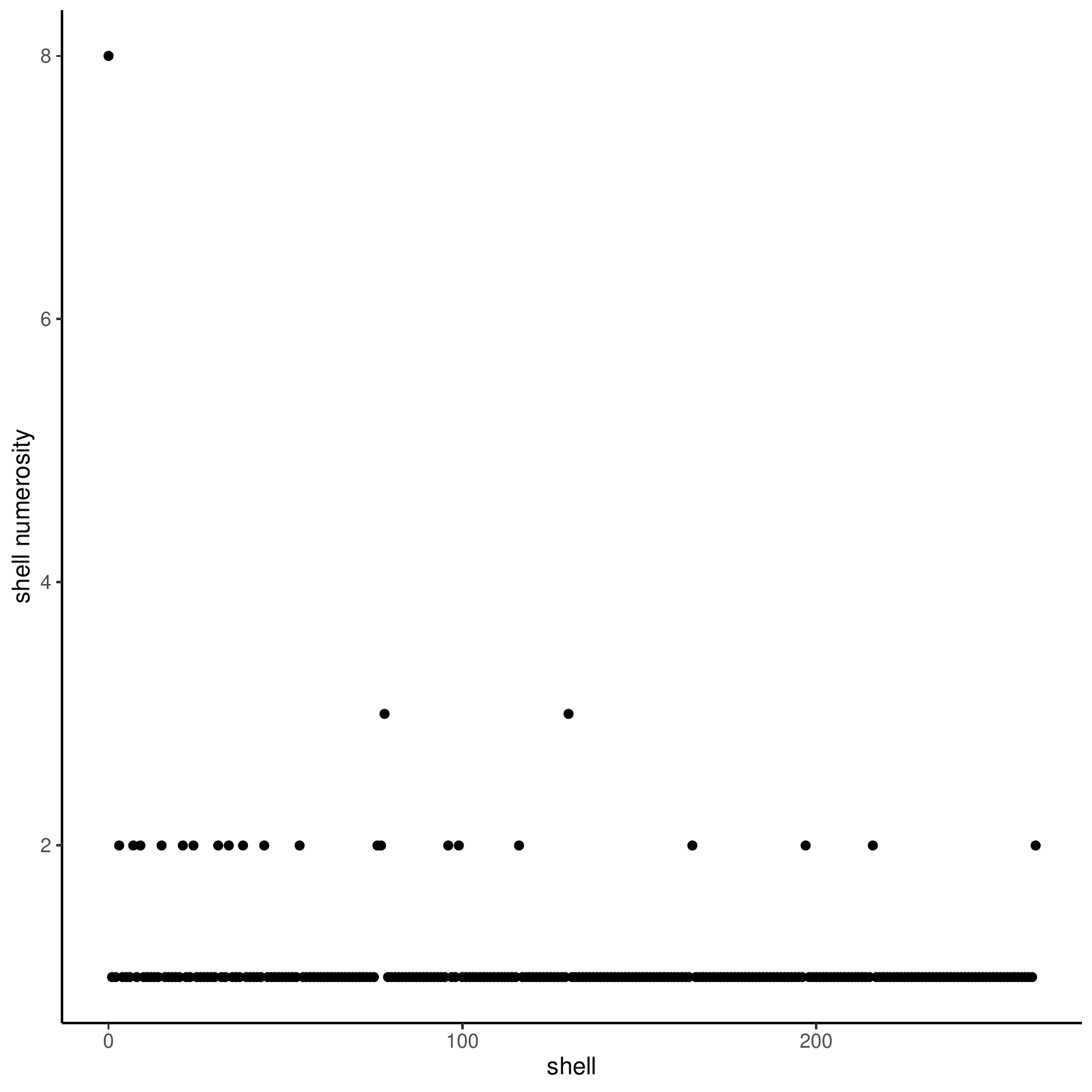}}
\end{minipage}
\end{figure}

To inspect the $s$-cores of the cumulative network we report the distribution of the thresholds in Figure \ref{figure: s-core distr} and the numerosity of each shell in Figure \ref{figure: s-core shell}, with shells ordered from the most peripheral to the most central ones.

\begin{figure}
\centering
\caption{\label{figure: s-core partition} Regions inside (darker) and outside (lighter) the core induced by the 90\% quantile of $s$-coreness distribution.}
\makebox{\includegraphics[scale=0.4]{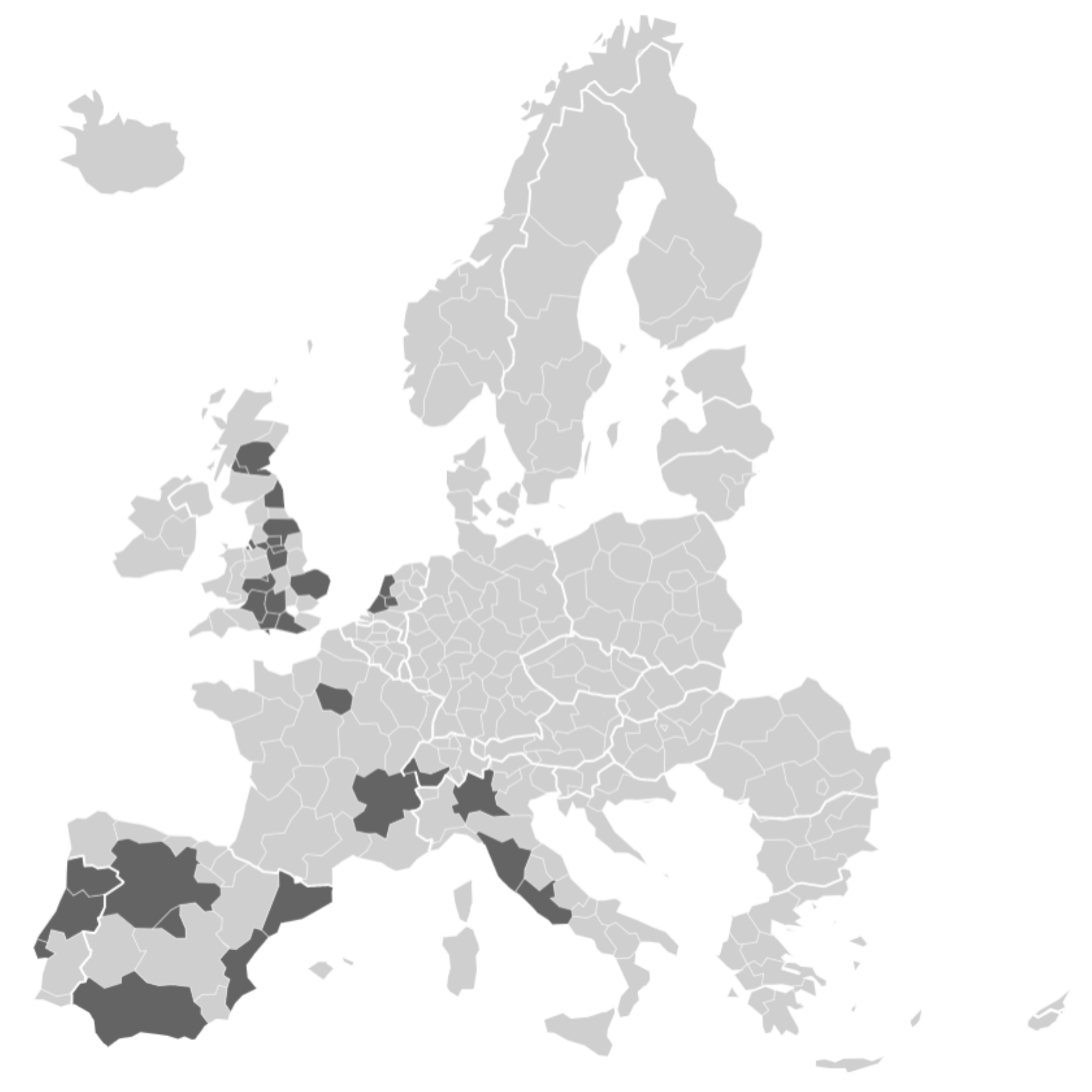}}
\end{figure}

By analyzing them, we recognize an almost continuous distribution of the thresholds (263 cores are present on top of 294 regions), with a few exceptionally central regions inducing the particularly high thresholds visible in Figure \ref{figure: s-core distr}.
We moreover notice almost every core differs from the previous by only one or two regions, given said continuity. We cannot therefore confidently find central networks of equally well connected regions, but we can define exceptional connected regions similarly to what was done with hubs and authorities in Section \ref{subsub: Hubs and Authorities Analysis}, by considering the most central core resulting from the top 10\% connected regions. This partition at the 90\% quantile on $s$-coreness distribution for the cumulative network can be visualized in Figure \ref{figure: s-core partition}.

We assess a general uniformity in time of this division by comparing it with the one induced by year-wise networks' $s$-coreness. By partitioning the nodes at the 90\% quantile for each year separately and computing the accuracy with respect to the partition induced by the cumulative network, to be interpreted as a congruence score between the partitions, we obtain scores between 84\% and 89\%.

We also notice a high correspondence (76.5\%) between the regions with top 10\% hubs and authorities scores and the top 10\% connected regions for $s$-coreness in the cumulative network. This suggests that the regions that act as strong providers and attractors of researchers are also the most connected in a central network of talent exchange.

Such results provide further evidence, together with the correlation studies of Section \ref{subsection: Correlation Study}, towards the validity of brain mobility theories where a high exchange of talents is key. We have indeed verified with different approaches how there is no clear distinction between sending regions and receiving regions in the network, while connectivity based approaches seem to fit the phenomenon best. Moreover, partitions induced by the general grade of mobility turn out to be meaningful and reflected in different regional characteristics. Said characterization in terms of mobility is further explored in Section \ref{section: Models}.

\section{Models}
\label{section: Models}

We now fully introduce the regional characteristics illustrated in Section \ref{subsection:coovariates} into our analysis. Section \ref{subsection: Network Model} explores the relationship from a network perspective, once again assessing a symmetry between the effects of a sender region and a receiving one. Since our results point to pure regional mobility being the key for understanding researcher's migrations, Section \ref{subsection: Mobility Model} models it with respect to selected regional characteristics.

All the results of this section are consistent with the ones obtained with national data, reported in Appendix \ref{Appendix C national network model} and Appendix \ref{Appendix C national mobility model}.

\subsection{Network Model}
\label{subsection: Network Model}

We set ourselves in a fully connected bidirectional network framework where nodes are NUTS2 regions and over every arc we have 12 different weights, each representing the total number of researchers migrating from one region another in the 12 years of the analysis.

\subsubsection{Additional Variables}

In order to build a model for the weights of such a bidirectional network, new variables able to capture the possible relationships between two different regions are needed.

We define the distance between two regions as the geographical distance between the respective regional centroids. Such a metric will be considered under logarithmic scale, to account for the order of magnitude of the spatial closeness between two regions.\\
As highlighted in Section \ref{subsection: Spatial Characterization}, the phenomenon is characterized by a strong component of movements inside a nation. This is modeled via a binary factor, same\_country, which indicates if an arc is defined over two regions which are part of the same country. Moreover we insert in our analysis another binary factor, same\_lan, which indicates if the dominant language of the two regions belongs to the same linguistic family (romance, germanic, slavic, \dots). A full description on how such linguistic families are defined is reported in Appendix \ref{Appendix A - language}.

\subsubsection{Modeling Choice}
The idea for our modeling choice is inherited from gravity models \cite{beine_practitioners_2016}, widely used for analysing migrations \cite{lewer_gravity_2008, van_bouwel_does_2010, botezat_physicians_2020}.

Our intention is to generalize their application through the use of non parametric regression, spline regression in particular, in order to allow for more complex relationships between a regions' characteristics and the flow of researchers between them. In particular, we do so through the use of generalized additive models \cite{gam}, since the structure of a gravity model can be expressed additively just by applying a logarithmic transformation. The only assumption of said model would hence be an additive structure over the effects of the regional characteristics. We apply a penalization on the second derivative of the smoothed terms, as the purpose of the model is explanatory on the phenomenon and we prefer to have interpretable effects even at the cost of a reduction in predictive power.

The time dependency of our observations, together with the time-wise heterogeneity explored in Section \ref{subsection: Temporal Characterization}, needs to be accounted for. 
Mixed effects models \cite{harrison_brief_2018, Jiang2022} are widely adopted as a way to account for hierarchical structures in the data, as they not only allow for factor-dependent effects but they also explicitly model the covariance structure within the observations, yielding for the assumption of \textit{i.i.d.} residuals to hold.
We introduce a random intercept for every year of the analysis, such a correction will indeed model the nested covariance structure and account for magnitude imbalances of the phenomenon across years. Additionally, we permit variables to have different effects across different years, leaving the possibility for an evolution of the phenomenon to be captured.

The complete form of our model is reported in Model (\ref{mod_network_full}).

\begin{eqnarray}\label{mod_network_full}
\log\left(flow_{year}(sender,receiver)+1\right) &=& f^{g, s}_{year}(gdp\_pc_{year}(sender)) +f^{g, r}_{year}(gdp\_pc_{year}(receiver)) +\nonumber \\
&+& f^{e, s}_{year}(edu_{year}(sender)) +f^{e, r}_{year}(edu_{year}(receiver)) + \nonumber \\
&+& f^{u, s}_{year}(uni_{year}(sender)) +f^{u, r}_{year}(uni_{year}(receiver)) +\nonumber \\
&+& f^{t, s}_{year}(TED_{year}(sender)) +f^{t, r}_{year}(TED_{year}(receiver)) +\nonumber \\
&+& g(\log dist(sender, receiver)) + \\
&+&\beta_l*same\_lan(sender, receiver)+\nonumber \\
&+& \beta_c*same\_country(sender, receiver)+\nonumber \\
&+& \alpha_{year} +
\varepsilon_{year}^{(sender, receiver)}\nonumber\\
\alpha_{year} \overset{\textit{iid}}{\sim} \mathcal{N}(0, \sigma), \ \varepsilon_{year}^{(sender, receiver)}\ \textit{iid}; &&\forall\, sender,\ receiver \in regions\, \forall\ year \in [2009, 2020]\nonumber
\end{eqnarray}

In the model $gdp\_pc$, $edu$, $uni$, and $TED$ represent the regional characteristic, year-wise and region-wise, introduced in Sections \ref{var gdp}, \ref{var edu}, \ref{var uni} and \ref{var ted} respectively.\\
The reason for having a +1 inside the logarithm is to make it such that empty arcs have a weight of zero. No parametric assumption on the distribution of the errors is made, tests are conducted following non parametric theory and detailed as they are used.

\subsubsection{Final Network Model}
\label{subsub: Network GAM final model}
\begin{figure}
\centering
\caption{\label{final gam network}Final network model's smooth components, University Score for sending (A) and receiving (B) region, TED for sending (C) and receiving (D) region and distance between the two regions in log scale (E).}
\makebox{\includegraphics[scale=0.45]{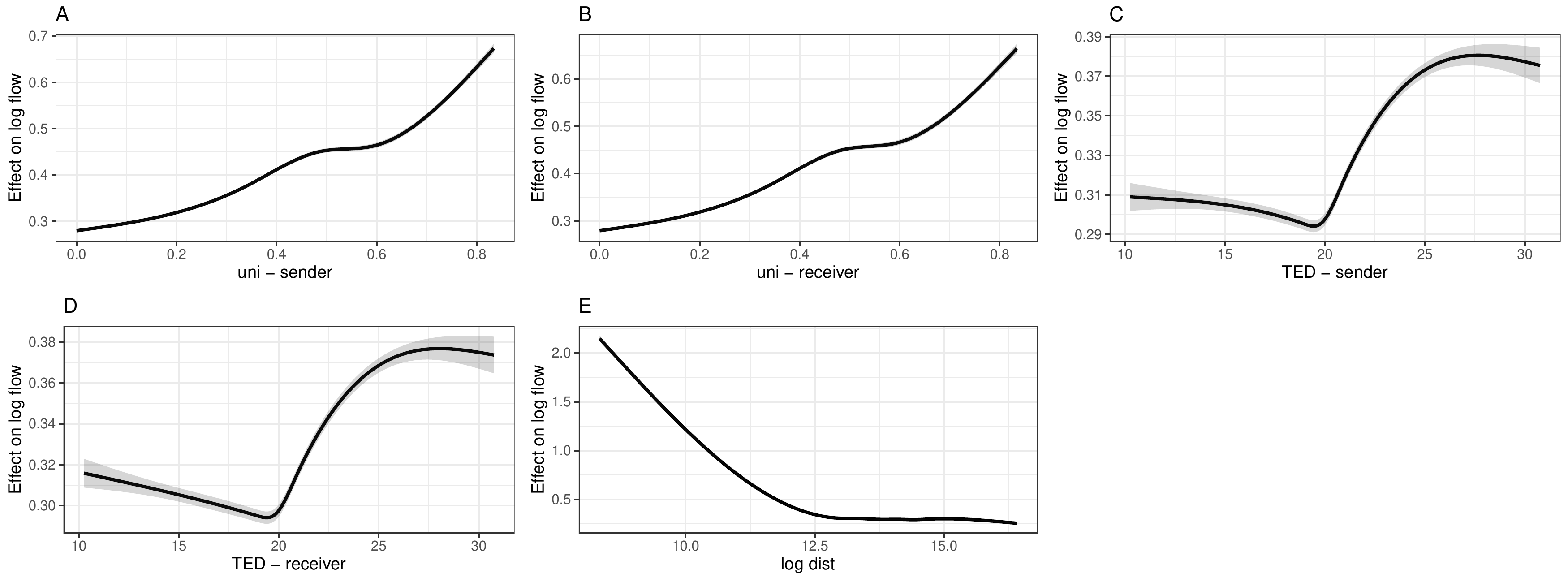}}
\end{figure}

Multiple models are possible, we propose one of the better performing (R-squared = 0.495) and more insightful on the phenomenon. 
Its general structure is reported in Model (\ref{mod_network_final}) and a visualization of its smoothed effects is displayed in Figure \ref{final gam network}.

\begin{eqnarray}\label{mod_network_final}
\log\left(flow_{year}(sender,receiver)+1\right) &=&  \overline{f}^{u, s}(uni_{year}(sender)) +\overline{f}^{u, r}(uni_{year}(receiver)) +\nonumber \\
&+& \overline{f}^{t, s}(TED_{year}(sender)) +\overline{f}^{t, r}(TED_{year}(receiver)) +\nonumber \\
&+& \overline{g}(\log dist(sender, receiver)) + \\
&+& \overline{\beta}_c*same\_country(sender, receiver)+\nonumber \\
&+& \overline{\alpha}_{year} +
\overline{\varepsilon}_{year}^{(sender, receiver)}\nonumber\\
\overline{\alpha}_{year} \overset{\textit{iid}}{\sim} \mathcal{N}(0, \overline{\sigma}), \ \overline{\varepsilon}_{year}^{(sender, receiver)}\ \textit{iid}; &&\forall\, sender,\ receiver \in regions\, \forall\ year \in [2009, 2020]\nonumber
\end{eqnarray}

Such a model is based on University Score and TED as regional characteristics for both senders and receivers, with the same effects across all years, while considering the logarithm of the distance between regions and the intra-national movement factor as interaction components between regions. Validation results and comparisons with alternative models are reported in Appendix \ref{Appendix B network model selection and robustness}.

It is straightforward to notice how the academic prestige, modeled by University Score, of both the sending and the receiving region has a positive impact on the amount of researchers traversing the edge, while the amount of investments, modeled by TED, has an almost step-like effect, again for both sending and receiving regions. Geographical closeness has a rapidly decreasing positive effect, and a positive impact is modeled by the intra-national component. Results on the robustness of this model are detailed in Appendix \ref{Appendix B network model selection and robustness}.

\subsubsection{Senders and Receivers Effects Symmetry}
\label{subsub symmetry}
A curious aspect of the model presented in Section \ref{subsub: Network GAM final model}, and present in all other plausible network models as well, is that sender and receiver regions seem to have interchangeable effects: by looking at Figure \ref{final gam network} one can notice that the function describing the impact of the regional characteristics of the sender region are exceptionally similar to those of the receiver region. Such an observation would imply that the characteristics that correlate with higher influxes of researchers are the same that, in the same way, correlate with higher losses of talents. 

To test this symmetry hypothesis we build Model (\ref{mod_symmetric}), where the regional characteristics of both sending and receiving regions are summed together and fed a single function.
\begin{eqnarray}
\label{mod_symmetric}
\log\left(flow_{year}(sender,receiver)+1\right)\qquad &=& f^{\star u}(uni_{year}(sender) + uni_{year}(receiver)) +\nonumber\\
&+& f^{\star t}_{year}(TED_{year}(sender) + TED_{year}(receiver)) +\nonumber\\
&+& g^{\star}(\log dist(sender, receiver)) +\\
&+& \beta^{\star}_c*same\_country(sender, receiver)+\nonumber\\
&+& \alpha^{\star}_{year} +
\varepsilon_{year}^{\star\ (sender, receiver)}\nonumber\\
\alpha^{\star}_{year} \overset{\textit{iid}}{\sim} \mathcal{N}(0, \sigma^{\star}), \ \varepsilon_{year}^{\star\ (sender, receiver)}\ \textit{iid}; && \forall\, sender,\ receiver \in regions\, \forall\ year \in [2009, 2020]\nonumber
\end{eqnarray}

If the intuition is true, a model which also has separate effects for senders' and receivers' characteristics should not be significantly better than this reduced model. A generalization of the structure of Model \ref{mod_network_final} is needed in order to frame our test as a variable selection one. For this reason we define Model \ref{mod_Asymmetric}, where the symmetric component of the model is explicitly kept into account.

\begin{eqnarray}
\label{mod_Asymmetric}
\log\left(flow_{year}(sender,receiver)+1\right)\qquad &=& \tilde{f}^{u}(uni_{year}(sender) + uni_{year}(receiver)) +\nonumber\\
&+&\tilde{f}^{u, s}(uni_{year}(sender)) + \tilde{f}^{u, r}(uni_{year}(receiver)) +\nonumber\\
&+& \tilde{f}^{t}_{year}(TED_{year}(sender) + TED_{year}(receiver)) +\nonumber\\
&+& \tilde{f}^{t, s}_{year}(TED_{year}(sender)) + \tilde{f}^{t, r}_{year}(TED_{year}(receiver)) +\\
&+& \tilde{g}(\log dist(sender, receiver)) +\nonumber\\
&+& \tilde{\beta}_c*same\_country(sender, receiver)+\nonumber\\
&+& \tilde{\alpha}_{year} +
\tilde{\varepsilon}_{year}^{(sender, receiver)}\nonumber\\
\tilde{\alpha}_{year} \overset{\textit{iid}}{\sim} \mathcal{N}(0, \tilde{\sigma}), \ \tilde{\varepsilon}^{\star\ (sender, receiver)}\ \textit{iid}; && \forall\, sender,\ receiver \in regions\, \forall\ year \in [2009, 2020]\nonumber
\end{eqnarray}

We hence formulate Test \ref{test symmetry}:
\begin{eqnarray}
\label{test symmetry}
&&H0: \mathcal{M} \in Model(\ref{mod_symmetric}) \qquad H1: \mathcal{M} \in Model(\ref{mod_Asymmetric}) \nonumber\\
&&Test \ statistic = max\{\text{F-statistics}[\tilde{f}^{u, s}, \tilde{f}^{u, r}, \tilde{f}^{t, s}, \tilde{f}^{u, r}]\}\\
&&Permutation \ scheme: permutation \ of \ residuals \ of \ the \ null\ model \nonumber
\end{eqnarray}
Test (\ref{test symmetry}) yields a p-value of 0.974, hence we do not have evidence to state that there is an asymmetric component to the phenomenon and Model \ref{mod_symmetric} is enough for describing it. 

This result, together with the analysis conducted in Section \ref{section: Mobility Analysis}, confirms the interpretation of the phenomenon being guided by pure mobility, rather then a net loss or gain of researchers. This is coherent with other proposed theories, like the notion of brain circulation \cite{yu_brain_2021}, and is consistent with similar observations done in other analysis \cite{urbinati_measuring_2021}.

\subsection{Mobility Model}
\label{subsection: Mobility Model}

Since the results of Section \ref{section: Mobility Analysis} and Section \ref{subsub symmetry} confirm a characterization of the phenomenon in terms of mobility, we model it with respect to regional characteristics.

We consider, for each region over each year, the grade of mobility, defined as logarithm of the sum of the total number of people entering plus the number of people leaving said region in the specific year.

\subsubsection{Modeling Choice}
The response variable to be modeled is hence the logarithm of the total mobility of each region, computed for all 12 years separately.
In order to allow for more complex relationships between regional characteristics and the grade of mobility, we again exploit generalized additive models theory \cite{gam}, fitting a cubic spline over each term with a penalization over the second derivative.

The regional datum presents two separate hierarchical structures: a time-dependent one, explored in Section \ref{subsection: Temporal Characterization} and a space-dependent one, explored in Section \ref{subsection: Spatial Characterization}. This means that regions belonging to the same country might have an additional level of dependency on top of the one induced by measurements referring to different years. Said dependencies are taken into account through the theory on mixed effects models \cite{Jiang2022}, using two sets of random intercepts: a year-wise one, which models the hierarchical structure induced by the repeated measure nature of our data, together with magnitude imbalances across years; a country-wise one, which models the hierarchical structure induced by regions belonging to different countries, together with possible country dependent effects which are not captured by the regional characteristics we select.

In order to allow for a possible evolution of the phenomenon, variables can have different effects across different years. The complete form is hence the one of Model \ref{mod_mob_full}.

\begin{eqnarray}
\label{mod_mob_full}
\log\left(in\_flow_{year}(region) +out\_flow_{year}(region) +1\right)\qquad &=& f^{g}_{year}(gdp\_pc_{year}(region))+\nonumber\\
&+& f^{u}_{year}(uni_{year}(region))+\nonumber\\
&+& f^{t}_{year}(TED_{year}(region))+\\
&+& f^{e}_{year}(edu_{year}(region))+\nonumber\\
&+& \alpha_{year} +\alpha_{country(region)} +
\varepsilon_{year}^{region}\nonumber\\
\alpha_{year} \overset{\textit{iid}}{\sim} \mathcal{N}(0, \sigma_{year}), \
\alpha_{country} \overset{\textit{iid}}{\sim} \mathcal{N}(0, \sigma_{country}), \ \varepsilon_{year}^{region}\ \textit{iid}; && \forall\, region \in regions\, \forall\ year \in [2009, 2020]\nonumber
\end{eqnarray}

With $gdp\_pc$, $uni$, $TED$ and $edu$ being the regional characteristic, year-wise and region-wise, introduced in Sections \ref{var gdp}, \ref{var edu}, \ref{var uni} and \ref{var ted} respectively.\\
The reason for having a +1 inside the logarithm is to make it such that regions where no researchers either leaving or entering it have a response of zero. No parametric assumption on the distribution of the errors is made so tests are conducted following non parametric theory.

\subsubsection{Final Mobility Model}
\label{subsub: final mobility model}

\begin{figure}
\centering
\caption{\label{final mobility model}Final Mobility Model's smooth components, University Score (A), TED (B) and Education Index (C).}
\makebox{\includegraphics[scale=0.45]{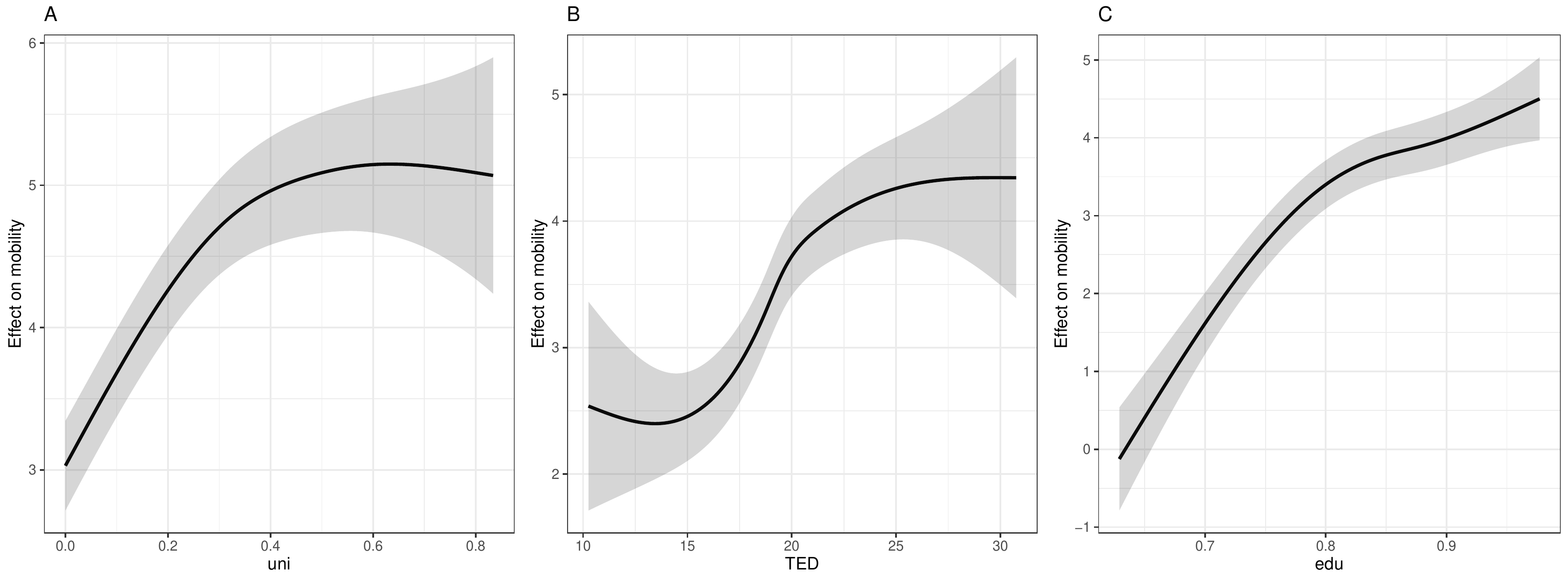}}
\end{figure}

From the model selection procedure multiple models are obtainable, we propose one of the better performing (R-squared = 0.686) and more insightful. Its general structure is reported in Model (\ref{mod_mob_final}) and a visualization of its smoothed effects is displayed in Figure \ref{final mobility model}.
\begin{eqnarray}
\label{mod_mob_final}
\log\left(in\_flow_{year}(region) +out\_flow_{year}(region) +1\right)\qquad &=& \overline{f}^{u}(uni_{year}(region))+\nonumber\\
&+& \overline{f}^{t}(TED_{year}(region))+\nonumber\\
&+& \overline{f}^{e}(edu_{year}(region))+\\
&+& \overline{\alpha}_{year} +\overline{\alpha}_{country(region)} +
\overline{\varepsilon}_{year}^{region}\nonumber\\
\overline{\alpha}_{year} \overset{\textit{iid}}{\sim} \mathcal{N}(0, \overline{\sigma}_{year}), \
\overline{\alpha}_{country} \overset{\textit{iid}}{\sim} \mathcal{N}(0, \overline{\sigma}_{country}), \ \overline{\varepsilon}_{year}^{region}\ \textit{iid}; && \forall\, region \in regions\ \forall\ years \in [2009, 2020]\nonumber
\end{eqnarray}
Such a model exploits University Score, TED and Education Index, having the same effects across all years. Validation results and comparisons with alternative models are reported in Appendix \ref{Appendix B mobility model selection and robustness}.

From the visualization of Figure \ref{final mobility model} it is possible to notice how the academic prestige represented by University Score, which reaches a plateau for very prestigious regions, has a positive correlation with mobility. Similarly to the network model of Section \ref{subsub: Network GAM final model}, the order of magnitude of investments, modeled by TED, behaves in a step-like fashion and ultimately the schooling level of the population, modeled by Education Index, has a positive correlation with mobility. Such results are robust with respect to possible outlying observations and leverage points, as detailed in Appendix \ref{Appendix B mobility model selection and robustness}.

This model, albeit not being proof of any causal link, still outlines patterns that should be further investigated. If causalities were to be assessed, policy makers would have a clear strategy for improving a regional researchers' mobility by targeting their efforts on specific aspects of their regions.
  
\section{Conclusions}
\label{section: conclusions}
\color{black}
In this work we present ORCID as a valuable source for tackling the researchers' migration phenomenon, circumventing the problem posed by the lack of official data. Well aware of its possible flaws we try to address valid concerns over its reliability by focusing the time-space horizon of the analysis. We extract a dataset covering NUTS2 regions of 31 European countries, over the time horizon of 2009-2020, on which two fundamental questions are raised: is the phenomenon of researcher's migrations to be interpreted under the traditional lens of brain drain or is there a more suitable characterization? What directions should a policy makers explore in order to improve the research environment of a region?

By conducting a preliminary analysis a year-wise heterogeneity is assessed. It is mostly driven by magnitude inflation on the overall number of migrations and it is usually not reflected in the phenomenon nature. By using community detection techniques we provide a general characterization of the migration network, highlighting both its overall uniform connectiveness but also the presence of a non-trascurable component of migrations happening inside national borders.

The opposition between the traditional interpretation of brain drain and the more recent interpretations is addressed, by looking for a suitable characterization for the researchers' migration phenomenon.\\
Through Spearman correlation testing, we assess a strong correlation between the incoming and outgoing flows of researchers in each region, highlighting how, in each region, the loss of talent is actually balanced by a gain of new incoming researchers and viceversa.
Analysing the network from a hubs and authorities perspective, coupled with the technique of $s$-coreness, we suspect the researchers' migration phenomenon to be effectively characterizable in terms of sheer mobility. We indeed find a strong correspondence between the biggest attractors and providers of talent, together with them being also the regions defining the most connected portion of the researchers' mobility network.\\
These results show how, in this setting, regions are not divisible between net providers and attractors, but a distinction can be made between those which are active and well inserted in the skill exchange network and those which are peripheral, neither attracting nor exporting a significant amount of talent.

To further explore the validity of this interpretation, we assess if partitions of regions based on mobility are reflected in those regions' characteristics. This is performed through a multi-aspect ANOVA test based on non parametric combinations theory, which detects differences in both location and scales of the two groups simultaneously. It is indeed the case that regional characteristics behave differently, particularly in their mean value, among induced groups, further confirming the meaningfulness of a mobility based interpretation.

Upon these findings we try to model researchers' migrations under a network perspective, generalizing gravity models. We obtain a model where university prestige and the amount of regional investments of both sender and receiving regions have a positive correlation with the amount of researchers traveling between the two. A symmetry between senders and receivers effects is explored and confirmed: it appears that the same qualities that correlate with attractive powers correlate in the same way with a higher loss of talent, while it is not the case that researchers migrate from low offering regions to higher offering ones.\\
This result, combined with the previous findings, supports the most recent trends in literature which revisit the brain drain interpretation of the skills migration phenomenon, at least in the space domain of Europe. We propose the framework of "brain mobility", where the focus is put on regions' connectiveness and ability to be inserted into the global research network.
As mobility is identified to be the focal point of the analysis, we further explore how it can be characterized in terms of regional predictors. We do so by proposing a model explaining the total number of researchers entering plus leaving a region as a function of university prestige, investments and education level of a region, all three showing a positive relationship with the mobility of said region.

We believe that such correlations, in order to be reliably identified as investments strategy for policy makers, should be further explored to assess possible causality links. We also believe that further research, the use of different statistical techniques, a more informed choice for the regional characterization or the analysis of different aspects of mobility, could lead to significant improvements on the current understanding of the topic.\\
Furthermore, as ORCID public registers become richer and richer, it would be feasible to expand the time-space domain of the analysis, possibly analysing the phenomenon outside of the boundaries of Europe. From ORCID data it is also possible to infer the stage of researcher's careers, such an information could be included in further analysis as a fragmentation of the sample could paint a different picture of this complex, but fascinating, phenomenon.



\bibliographystyle{plain}
\bibliography{reference}


\newpage

\appendix

\section{Data Extraction and Pre-Processing Details}
\label{Appendix A}
This appendix is devoted to go into details about the data extraction procedure and our choices for missing data imputation.

We underline how such pre-processing is a necessary step, as data availability at the regional level is often unavailable, incomplete or uses inconsistent definitions of regional boundaries. We highlight the importance of refining and uniforming such data collection processes, as it would be of great utility for performing in depth analysis no longer constrained by national borders, allowing at the same time the same degree of robustness and precision by relying on higher quality data.

\subsection{TED Data Construction}
\label{appendix a: ted}
Tenders Electronic Daily public procurement notices, whose source is reported in Table \ref{ted_data}, offer for every year a table in which informations on each procurement are available. In particular we extract, for every procurement notice: the year of the notice, the postal code associated to the notice, the country code of the procurement, the NUTS, if reported, three voices for the value of the procurement (a primary one and two secondary ones which are used if the primary is not reported) and the informations of an eventual non award.

Firstly we filter the procurements which have not been awarded, we then intervene to:
\begin{enumerate}
    \item Assign the value of the procurement, filling the missing values in the primary one (VALUE\_EURO) with the two secondaries (VALUE\_EURO\_FIN\_1 and VALUE\_EURO\_FIN\_2).
    \item Assing each procurement to a region, we do so by means of the reported NUTS, where available and if reported at the desired granularity. Otherwise we use Postal Code to infer the NUTS2 region, following the same procedure reported in Section \ref{subsub: ORCID Data extraction}. 
\end{enumerate}

Since it is not uncommon for data not to
be input incorrectly or missing (as reported in data documentation), not all notices have been successfully mapped into the corresponding region and not all procurement values are available. We filter the table in order to consider only the procurements for which both these informations have been obtained.

It now suffices to sum, for every region in each year, the values of the appropriate procurements. Such values are subject to an exceptionally high variability and are not robust. On top of having to deal with a partial information induced by data unavailability and possible manual errors in the reporting process, there is the presence of notices with unrealistically high values (in the order of magnitude of thousands of billions of euros). We do not remove these observations, as such notices constitute a marginal part of the dataset, with only 0.3\% of the procurement values exceeding 100 million euros and 0.001\% being in the unrealistically thousands of billions. Instead we opt for a scaling of such values which helps in the robustness of the procedure and we consistently check if such observation skew our findings, as reported in Appendix \ref{Appendix B network model selection and robustness} and Appendix \ref{Appendix B mobility model selection and robustness}, which confirm the stability of our results.

By applying the logarithm of the sum of values, we not only offer a more meaningful variable with which to conduct the analysis, but we also curb the outlyingness of such exceptional procurement values.
It also drastically reduces the sample variability of the obtained indicator for each region, such variability reduction makes the imputation process much more stable.
In fact, for what concerns the investments magnitude, we assumed consistency of a region with respect to the correspondent country and year, and we imputed with Amelia the values equal to zero treating them as missing data. The imputations percentage with respect to total amount of data is $12.15\%$.

\begin{table}[t]
\begin{center}
\caption{Data Sources for Tenders Electronic Daily}
\label{ted_data}    
\begin{tabular}{@{}p{0.25\linewidth} p{0.65\linewidth}@{}}
\toprule
 Object &  Link\\
\midrule
TED public procurement notices  & \href{https://data.europa.eu/data/datasets/ted-csv/?locale=en}{\textit{https://data.europa.eu/data/datasets/ted-csv/?locale=en}}\\
\bottomrule
\end{tabular}
\end{center}

\end{table}

\subsection{GDP Data Construction}
\label{appendix a: gdp}

As stated in Section \ref{var gdp}, we used multiple sources in order to extract the raw data for GDP per capita, a summary of such sources and the relative links is reported in Table \ref{gdp_data}.

\begin{table}[t]
\begin{center}
\caption{Data Sources for GDP per capita}
\label{gdp_data}
\begin{tabular}{@{}p{0.25\linewidth} p{0.65\linewidth}@{}}
\toprule
Object & Link\\
\midrule
National GDP\_pc values with PPP correction & \href{https://data.worldbank.org/indicator/NY.GDP.PCAP.PP.CD}{\textit{https://data.worldbank.org/indicator/ NY.GDP.PCAP.PP.CD}}\\
Regional GDP\_pc (most countries) & \href{https://ec.europa.eu/eurostat/databrowser/view/tgs00005/default/table?lang=en}{\textit{https://ec.europa.eu/eurostat/databrowser/view/ tgs00005/default/table?lang=en}}\\
Regional (TL2) GDP\_pc for CH & \href{https://stats.oecd.org/Index.aspx?DataSetCode=REGION_ECONOM}{\textit{https://stats.oecd.org/Index.aspx?DataSetCode=REGION \_ECONOM}}\\
Regional (NUTS2) GDP per NO & \href{https://ec.europa.eu/eurostat/databrowser/view/TGS00003/default/table?lang=en}{\textit{https://ec.europa.eu/eurostat/databrowser/view/ TGS00003/default/table?lang=en}}\\
Regional (TL3) GDP for UK & \href{https://stats.oecd.org/Index.aspx?DataSetCode=REGION_ECONOM}{\textit{https://stats.oecd.org/Index.aspx?DataSetCode=REGION \_ECONOM}}\\
Regional (TL3) population for UK and NO & \href{https://stats.oecd.org/Index.aspx?DataSetCode=REGION_DEMOGR}{\textit{https://stats.oecd.org/Index.aspx?DataSetCode=REGION\_DE MOGR}}\\
Population of \textit{NO0B} & \href{https://www.ssb.no/en/befolkning/folketall/statistikk/befolkningen-pa-svalbard}{\textit{https://www.ssb.no/en/befolkning/folketall/statistikk/ befolkningen-pa-svalbard
}}\\
\bottomrule
\end{tabular}
\end{center}
\end{table}

Eurostat directly provides most country's regional GDP per capita, with most regions' values being reported for the whole time horizon. Notable exceptions are however present, namely:
\begin{enumerate}
    \item \textbf{Iceland}: for which, being the whole country a single NUTS2, no pre-processing is needed and we just use the national datum.
    \item \textbf{Switzerland}: for which OECD regional data was used being the regional TL2 corresponds to the NUTS2 classification in Switzerland.
    \item \textbf{Norway}: for which the changes to NUTS2 borders and the introduction of \textit{NO0B} as a new region required GDP per capita values to be hand computed from GDP data and population data. While GDP data were available for the required granularity in Eurostat, population data were aggregated from OECD regional data, being the TL3 classification corresponding to NUTS3 in Norway.
    \item \textbf{United Kingdom}: for which the changes to NUTS2 borders required values to be hand computed from GDP and population data, similarly to Norway this procedure is done on top of regional TL3 data, being corresponding to NUTS3 in the UK.
\end{enumerate}

For Norway and United Kingdom smaller NUTS3 regions were mapped into the corresponding NUTS2 regions according to the 2021 classification, then the respective GDP and population values were summed region-wise, obtaining raw GDP and population values for NUTS2 regions. Data for London, being itself technically a NUTS1 made of different NUTS2, was further aggregated. Population data for the region Jan
Mayen and Svalbard, \textit{NO0B}, was imputed according to the data for the population of Svalbard alone since, to the best of our knowledge, Mayen does not have any stable residents. The population estimation procedure for Norway required strong levels of imputing, being regional data for population available, for many regions, only from 2014 onward.

After having manually computed the per capita values of GDP for the UK and NO, we have a mostly fully table for all regions. Such table however is not consistent in unit of measure and Purchasing Power corrections, for this reason we use the national data extracted from World Bank, being both complete and PPP-adjusted, as benchmark. In all our regional values, we have, indeed, a national one as well. We compute the year-wise country-wise correction factors as:
\[
\alpha(country, year) = \frac{GDP\_pc^{World\ Bank}(country, year)}{GDP\_pc^{table}(country, year)}
\]

And we apply them to the regional values as:
\[
GDP\_pc^{corrected}(region, year) = \alpha(country(region), year)*GDP\_pc^{table}(region, year)
\]

Hence forcing the regional observations to be coherent with the reported national values. Being changes in currency and purchasing power corrections both multiplicative factors and nation-wise applied, this procedure does not introduce significant biases in the analysis.

Few observations ($8.33\%$) are still missing and are imputed with Amleia after pre-processing.

\subsection{Education Index Data Construction}
\label{appendix a: edu index}

Sources for the tables used in the extraction and pre-processing of the regional values for the Education Index which we use in the analysis are reported in Table \ref{edu_data}. The regional values for Education Index are however not reported at the required level of granularity for all the nations of the analysis, so two levels of corrections were applied.
\begin{table}[t]
\begin{center}
\caption{Data Sources for Education Index}
\label{edu_data}
\begin{tabular}{@{}p{0.25\linewidth} p{0.65\linewidth}@{}}
\toprule
 Object & Link\\
\midrule
Education Index & \href{https://globaldatalab.org/shdi/table/edindex/}{\textit{https://globaldatalab.org/shdi/table/edindex/}} \\
Education attainment & \href{https://ec.europa.eu/eurostat/databrowser/view/EDAT_LFSE_04/default/table?}{\textit{https://ec.europa.eu/eurostat/databrowser/view/ EDAT\_LFSE\_04/default/table?}}\\
\bottomrule
\end{tabular}
\end{center}
\end{table}

For the nations where the reported regional granularity was too fine (often at the NUTS3 level), namely Croatia, France, Ireland, Latvia, Lithuania, Norway, Romania and Slovenia, some regions had to be aggregated in order to coincide with the NUTS2 definition. Such aggregation is done by simply computing the average Education Index across the regions being aggregated into one, being such smaller regions often comparable in size.\\
For the nations where the reported regional granularity was not fine enough, or was inconsistent with the NUTS2 definition, with the notable examples of Germany, Hungary and United Kingdom, together with few regions from various other nations, we exploit the dependency between the Education Index and the Education Attainment, in percentage, of the population between 25 and 64 years of age. We do so by fitting a Spline regression model with the NUTS2 regions whose Education Index is available, reported in Model (\ref{mod_edu_in}).
\begin{eqnarray}
\label{mod_edu_in}
&& Education\_Index_{year}(region) = f_{year}(Education\_attainment_{year}(region))\\
&& \forall\, region \in regions\ \forall\ years \in [2009, 2020]\nonumber
\end{eqnarray}

Model (\ref{mod_edu_in}), unlike the other models being presented, is selected not for interpretability but for sheer predictive power (R-squared = 0.752). Splines are penalized in order to prevent overfitting and increase result's reliability.

Model \ref{mod_edu_in} is ultimately used to predict the Education Index for most missing values as some measurements were not available, notably the ones from 2020.  

Education Index is missing for some regions after pre-processing, specifically $9.01\%$ of them, these values are imputed through Amelia by exploiting dependencies of regions in the same country and same year.

\subsection{University Score}
\label{Appendix  A - uniscore}
University Score is computed on top of the top 500 University Ranking proposed by QS, such rankings are accessible either directly on QS's official website and similar sources, for the most recent years, or on past iterations of the site, stored in the Internet Archives. A complete list of links is reported in Table \ref{uniscore_data}.
\begin{table}[t]
\begin{center}
\caption{Data Sources for University Ranking}
\label{uniscore_data}
\begin{tabular}{@{}p{0.15\linewidth} p{0.75\linewidth}@{}}
\toprule
 Object & Link\\
\midrule
2020 rankings & \href{https://www.topuniversities.com/university-rankings/world-university-rankings/2020}{\textit{https://www.topuniversities.com/university-rankings/world-university-rankings/2020}} \\
2019 rankings & \href{https://www.universityrankings.ch/en/results/QS/2019}{\textit{https://www.universityrankings.ch/en/results/QS/2019}}\\
2018 rankings & \href{https://www.universityrankings.ch/en/results/QS/2018}{\textit{https://www.universityrankings.ch/en/results/QS/2018}}\\
2017 rankings & \href{https://www.universityrankings.ch/en/results/QS/2017}{\textit{https://www.universityrankings.ch/en/results/QS/20172}}\\
2016 rankings & \href{https://www.universityrankings.ch/en/results/QS/2016}{\textit{https://www.universityrankings.ch/en/results/QS/2016}}\\
2015 rankings & \href{https://web.archive.org/web/20160310122908/https://www.topuniversities.com/university-rankings/world-university-rankings/2015}{\textit{https://web.archive.org/web/20160310122908/https://www.topuniversities.com/university-rankings/world-university-rankings/2015}}\\
2014 rankings & \href{https://web.archive.org/web/20160531142050/http://www.topuniversities.com/university-rankings/world-university-rankings/2014}{\textit{https://web.archive.org/web/20160531142050/http://www.topuniversities.com/university-rankings/world-university-rankings/2014}}\\
2013 rankings & \href{https://web.archive.org/web/20151025234030/http://www.topuniversities.com/university-rankings/world-university-rankings/2013#sorting=rank+region=+country=+faculty=+stars=false+search=}{\textit{https://web.archive.org/web/20151025234030/http://www.topuniversities.com/university-rankings/world-university-rankings/2013}}\\
2012 rankings & \href{https://web.archive.org/web/20130225025744/http://www.topuniversities.com/university-rankings/world-university-rankings/2012}{\textit{https://web.archive.org/web/20151025234030/http://www.topuniversities.com/university-rankings/world-university-rankings/2013}}\\
2011 rankings & \href{https://web.archive.org/web/20120429094102/http://www.topuniversities.com/university-rankings/world-university-rankings/2011}{\textit{https://web.archive.org/web/20120429094102/http://www.topuniversities.com/university-rankings/world-university-rankings/2011}}\\
2010 rankings & \href{https://web.archive.org/web/20110717074903/http://www.topuniversities.com/university-rankings/world-university-rankings/2010}{\textit{https://web.archive.org/web/20110717074903/http://www.topuniversities.com/university-rankings/world-university-rankings/2010}}\\
2009 rankings & \href{https://web.archive.org/web/20100102005901/http://topuniversities.com/university-rankings/world-university-rankings/2009/results}{\textit{https://web.archive.org/web/20100102005901/http://topuniversities.com/university-rankings/world-university-rankings/2009/results}}\\
\bottomrule
\end{tabular}
\end{center}
\end{table}

\subsection{Language Families}
\label{Appendix A - language}
The models studied in Section \ref{subsection: Network Model} make use of one additional variable, which try to model the easiness of linguistic adaptation from the sending to the receiving region.

We define for this purpouse 6 linguistic families, defined to indicate a similar structure in the dominant language of a region. Such families are:
\begin{enumerate}
    \item \textbf{Germanic}: spanning the countries of Austria, Germany, Denmark, Ireland, Iceland, Luxembourg, Netherlands, Norway, Sweden and United Kingdom, together with the Flemish Region of Belgium and Switzerland (with the exception of Ticino and Région lémanique).
    \item \textbf{Slavic}: where the countries of Bulgaria, Czechia, Croatia, Poland, Slovenia and Slovakia are mapped into.
    \item \textbf{Romance}: spanning the countries of Spain, France, Italy, Malta, Portugal and Romania, togheter with the belgique regions of Wallonia and Brussels and the swiss territories of Ticino and Région lémanique.
    \item \textbf{Uralic}: where the countries of Estonia, Finland and Hungary are mapped into.
    \item \textbf{Baltic}: language family that include Latvia and Lithuania.
    \item \textbf{Greek}: official language of Greece and Cyprus.
\end{enumerate}

In the restricted framework of researchers' migrations, however, it should be taken into account that English is an almost universally spoken language. Such observation may significantly impact in reducing linguistic barriers, and possibly explain the statistical insignificance of such a partition in the model proposed in Section \ref{subsub: Network GAM final model}.

\section{NUTS2 Models Selection and Robustness to Normalizations}
\label{NUTS2 models}
In this Appendix we detail the model selection procedure, the robustness of the proposed models and considerations about other interpretations of the phenomenon.

Appendix \ref{Appendix B selection and robustness} explain the model selection procedure for the network model of Section \ref{subsub: Network GAM final model} and for the mobility model of Section \ref{subsub: final mobility model} respectively, together with their robustness.
Appendix \ref{Appendix B normalized pop} shows network and mobility model that take into account regional populations, while Appendix \ref{Appendix B normalized res} contains the analysis while taking into account a regions' population of researchers.

\subsection{Validation and Robustness}
\label{Appendix B selection and robustness}

Appendix \ref{Appendix B network model selection and robustness} details the network model of Section \ref{subsub: Network GAM final model}, while Appendix \ref{Appendix B mobility model selection and robustness} details the mobility model of Section \ref{subsub: final mobility model}.

\subsubsection{Network Models: Validation and Robustness}
\label{Appendix B network model selection and robustness}

The performance metric of choice is the R-squared of the models, as with \numprint{1033704} unique responses we deem the risk of overfitting to be relatively low.

The goal of the models proposed in our analysis is to provide insights on the migration phenomenon, more than raw predictive accuracy, for this reason variables are chosen mostly for the smoothness of their fit and the resulting interpretability of the model, as long as no major drops in performance occurred.

We now mention two examples of alternative models, one with time-heterogeneous effects and another with all available regional characteristics, showing how there is little improvement in the performance metric with respect to the reduced model proposed in Section \ref{subsub: Network GAM final model}.

In particular the version of Model (\ref{mod_network_final}) (R-squared of 0.496) with year-wise effects for regional characteristics has a marginal increase of R-squared (+0.036), while using 44 additional smooth components, we hence decide to use the simpler model.
A model considering all regional characteristics, instead, has an R-squared of 0.503, again with only a marginal (+0.007) increase in performance at the cost of 8 additional smooth components.
We stress the fact that the symmetry described in Section \ref{subsub symmetry} is also evident the two models we just mentioned and in all other produced models.

Residual's distribution is heavily not gaussian, but still continuous. There are observations that, either in the residuals or in the covariates, could be labeled as outlying and we validate the robustness of the model by fitting the same formula over the subnetwork where such outlying nodes (regions) are removed. Notable examples of such regions are the city of London, $UKI0$, which is an outlier in terms of academic prestige and presents most of the outlying measurements for TED we explain in Appendix \ref{appendix a: ted}, and the portuguese regions of Norte and Centro, $PT11$ and $PT16$ respectively, which have exceptional spikes in mobility.\\
All models fitted on subnetworks show similar smoothed components and lead to the same interpretation as the original one. We hence deem the proposed model of Section \ref{subsub: Network GAM final model} to be robust and meaningful.

\subsubsection{Mobility Models: Validation and Robustness}
\label{Appendix B mobility model selection and robustness}

The performance metric of choice is the R-squared and model selection is performed in a backward fashion.
Since mobility models, compared to the network ones, are much smaller and more agile to fit, it is feasible to additionally formally compare them by permutational F-tests, computing the permutational distribution of the test statistic since gaussianity assumptions on the residuals are not verified. Such procedure indeed requires to fit multiple (in our case one thousand) different models fitted on pooled datasets where the residuals of the null (reduced) model have been permuted between observations.

The model proposed in Section \ref{subsub: final mobility model} has been selected for the balance between predictive performance (with an R-squared of 0.686) and interpretability, but alternative model definitions whit similar results and interpretations can be obtained. A model where regional characteristics are allowed to have different impacts in different years is not significantly better in terms of R-squared (being equal to 0.695) and the significance of such reduction is also tested through permutational F-test: with p-value equal to 1 we keep the time-invariant model.\\
A full model, using all available regional characteristics, is not significantly better either, it has an R-squared of 0.687, and the permutational F-test provides a p-value of 0.95, hence we keep the reduced model.

\subsection{Normalized Flows by Population}
\label{Appendix B normalized pop}

The sources for population data are reported in Table \ref{pop_data}. We extract for each region its population, and we analyse a normalized versions of the researchers flows. By normalization we mean to adjust the number of migrating researchers, either in the directed network or aggregated mobility setting, by instead considering its ratio over 100 thousand inhabitants of the region.

\begin{table}[t]
\begin{center}
\caption{Data Sources for Regional Population}
\label{pop_data}
\begin{tabular}{@{}p{0.25\linewidth} p{0.65\linewidth}@{}}
\toprule
Object & Link\\
\midrule
Regional (NUTS2) population for most countries & \href{https://ec.europa.eu/eurostat/databrowser/bookmark/894b774a-6868-4650-92e2-553948dfae2f?lang=en}{\textit{https://ec.europa.eu/eurostat/databrowser/bookmark/ 894b774a-6868-4650-92e2-553948dfae2f?lang=en}}\\
Regional (TL3) population for UK and NO & \href{https://stats.oecd.org/Index.aspx?DataSetCode=REGION_DEMOGR}{\textit{https://stats.oecd.org/Index.aspx?DataSetCode=REGION\_DE MOGR}}\\
Population of \textit{NO0B} & \href{https://www.ssb.no/en/befolkning/folketall/statistikk/befolkningen-pa-svalbard}{\textit{https://www.ssb.no/en/befolkning/folketall/statistikk/ befolkningen-pa-svalbard
}}\\
\bottomrule
\end{tabular}
\end{center}
\end{table}

\subsubsection{Mobility Models: Normalized Flows by Population}
\label{Appendix B mobility model normalized pop}

\begin{figure}
\centering
\caption{\label{figure: mobility model pop} Population normalized mobility models' smooth components. Including $UKI0$: University Score (A), TED (B), and Education Index (C). Excluding $UKI0$: University Score (D), TED (E) and Education Index (F).}
\makebox{\includegraphics[scale=0.45]{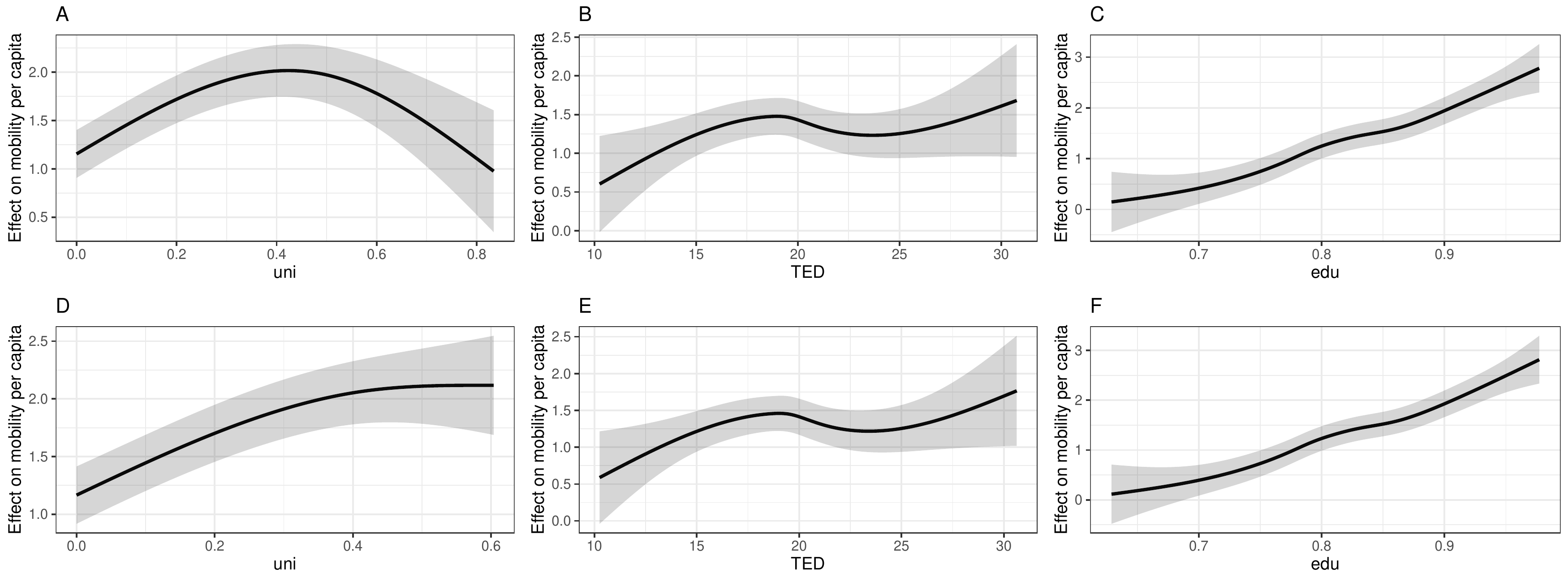}}
\end{figure}

As illustrated in Figure \ref{figure: mobility model pop}, after model selection, we notice a peculiar fitting for University Score, which is indeed caused by the city of London, $UKI0$, where its outlying nature in terms of the number of prestigious universities is highlighted. We fit the model removing such observation and what we notice is that the smooth components return to be similar to those of the model proposed in Section \ref{subsub: final mobility model}. It should be however noted that the analysis being conducted at the NUTS2 level (with the exception of London, which is indeed outlying), population sizes should be mostly comparable.

\subsubsection{Network Models: Normalized Flows by Population}
\label{Appendix B network model normalized pop}

When normalizing the flows in a network setting we use three available options: discounting the number of researchers travelling from a region to another with respect to the sending region's population, the receiving region's population and the sum of both.

When normalization is applied with respect to only one regional population, no significant model is produced, with models' R-squared being at most 0.200 and only 0.016 higher than those of a model which does not account for regional characteristics at all and is based purely of intercepts and geographic distance.

\begin{figure}
\centering
\caption{\label{figure: network model pop} Population normalized network models' smooth components: University Score for sending (A) and receiving (B) region, TED for sending (C) and receiving (D) region and distance between the two regions in log scale (E)..}
\makebox{\includegraphics[scale=0.45]{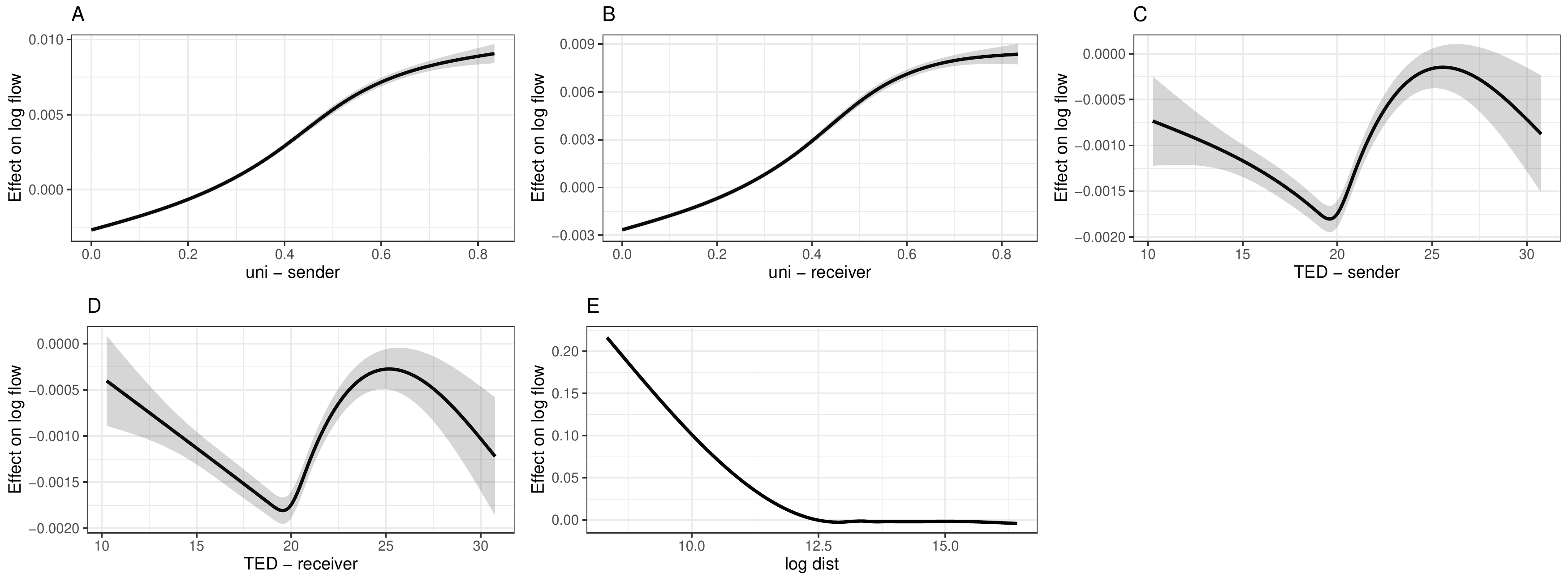}}
\end{figure}

A more meaningful model (R-squared equal to 0.429) is available when the normalization is performed using both regional population. A visualization of the smooth components is reported in Figure \ref{figure: network model pop}. Such visualization leads to similar interpretations as those obtained in Section \ref{subsub: Network GAM final model}.

\subsection{Normalized Flows by Number of Researchers}
\label{Appendix B normalized res}

From ORCID public data we extract, following a similar procedure to that described in Section \ref{subsub: ORCID Data extraction}, the number of researchers present in each region each year. We use this data to discount the migration flows, viewing them as the mobile portion of a regional research population.

\subsubsection{Mobility Models: Normalized Flows by Researchers}
\label{Appendix B mobility model normalized res}

The models that try and capture the portion of mobile researchers in each region are ultimately ineffective. We compare a full model, employing all regional characteristics, with a model which employs only the required random intercepts for countries and years. Such comparison is performed by means of an F-test, providing a p-value of 0.65, hence disproving the significance of the full model. Such weakness of the selected variables is also confirmed by the lack of predictive power, indeed while the intercept-only model has an R-squared of 0.365, a full model has an R-squared of 0.368, with only a marginal increase.

\subsubsection{Network Models: Normalized Flows by Researchers}
\label{Appendix B network model normalized res}

The correction being applied with respect to different regional research populations leads to different interpretations of the phenomenon. In all scenarios model selection is performed and all models are suboptimal in predictive power, with full models having only marginal (always less than 0.02) increases in R-squared with respect to those without regional characteristics. Model selection does not indicate clearly better performing models, we hence present those using the same variables as the one proposed in Section \ref{subsub: Network GAM final model} as they still are marginally better and provide an easy comparison.

\begin{figure}
\centering
\caption{\label{figure: res_net_send} Smooth components of the network model using sending region's researchers population. University Score for sending (A) and receiving (B) region, TED for sending (C) and receiving (D) region and distance between the two regions in log scale (E).}
\makebox{\includegraphics[scale=0.45]{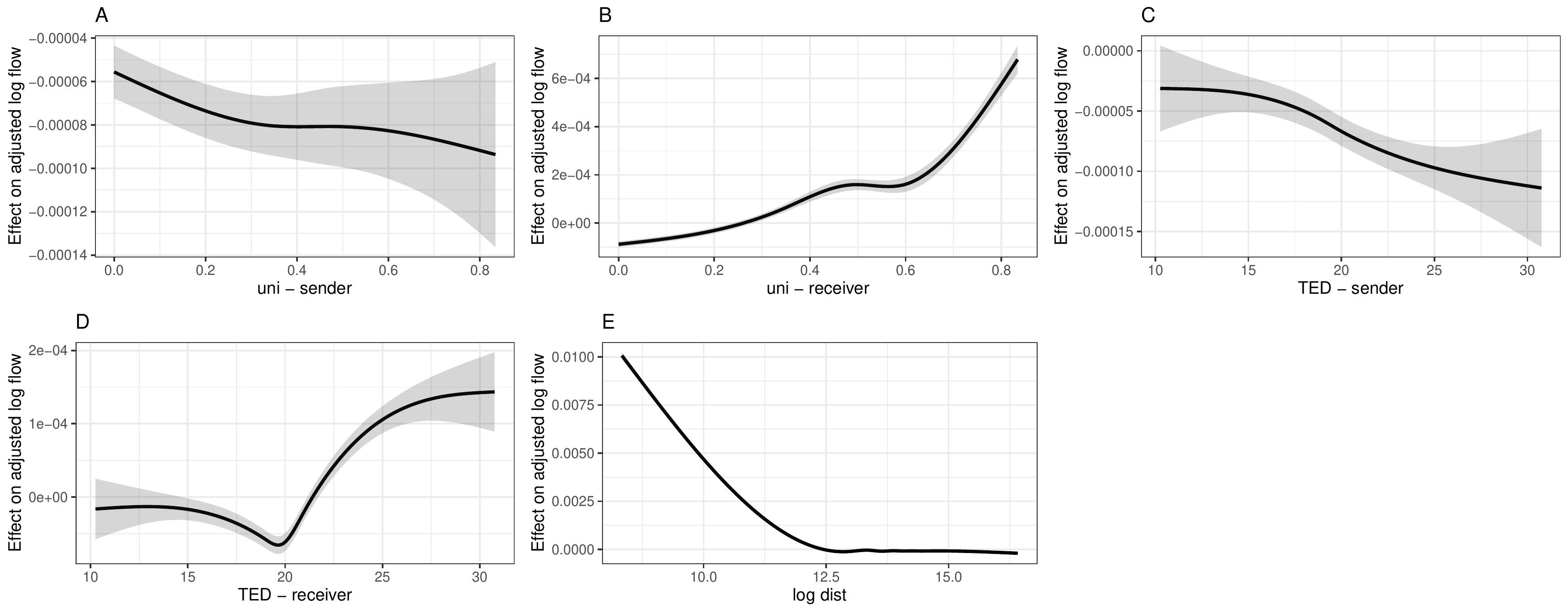}}
\end{figure}

When the arc weights are defined as the number of migrating researchers over the number of researchers in the sending region, we are substantially studying the probability that a researcher in the sending region migrates to the receiving one. The visualization of the smooth components of the proposed model for this scenario is reported in Figure \ref{figure: res_net_send}. We notice that the effects of the receiving region's characteristics are similar to those of the model presented in Section \ref{subsub: Network GAM final model}, with the effects of sending region's characteristics being 10 times smaller in magnitude. We are hence led to believe that the probability of a researcher deciding to migrate into another region is mostly driven by the destination's characteristics, with little impact being attributed to the region the researcher would be leaving.

\begin{figure}
\centering
\caption{\label{figure: res_net_rec} Smooth components of the network model using receiving region's researchers population. University Score for sending (A) and receiving (B) region, TED for sending (C) and receiving (D) region and distance between the two regions in log scale (E).}
\makebox{\includegraphics[scale=0.45]{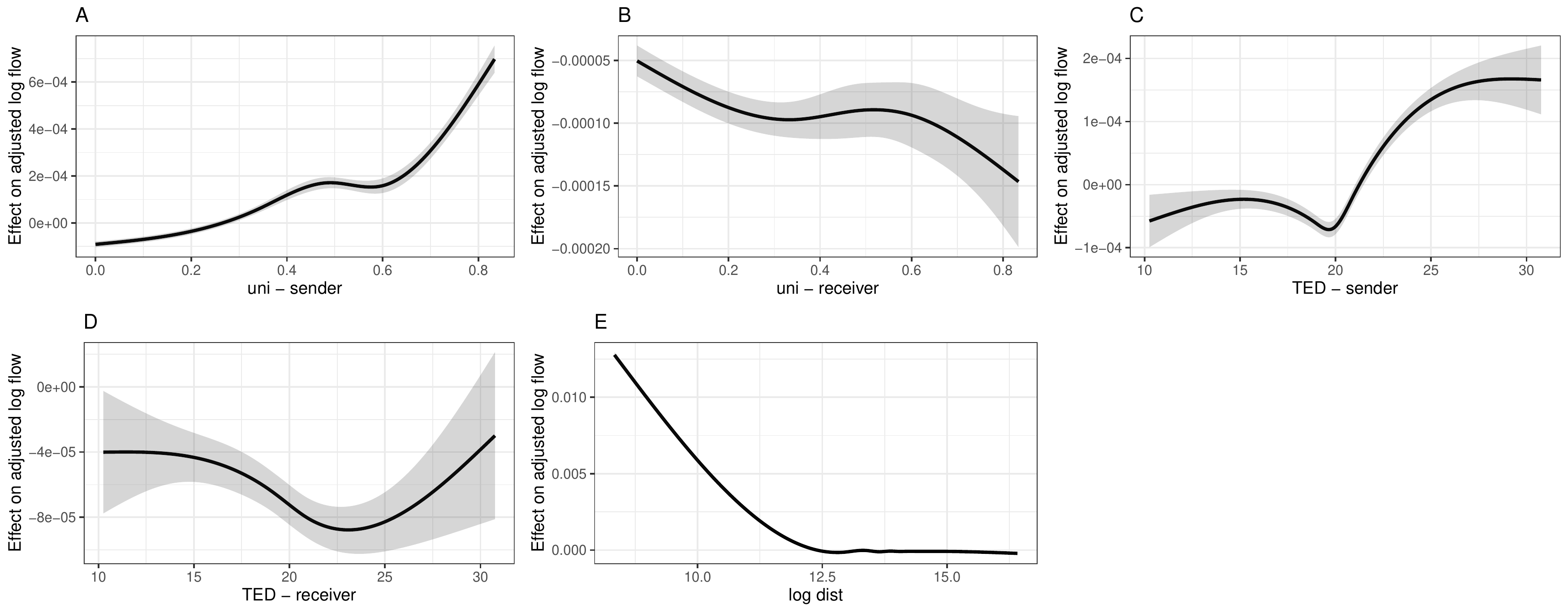}}
\end{figure}

Counter wise, when the arc weights are defined as the number of migrating researchers over the number of researchers in the receiving region, we are substantially studying the probability that a researcher in the receiving region has just migrated from the sending one. The visualization of the smooth components of the proposed model for this scenario is reported in Figure \ref{figure: res_net_rec}. As one can notice the fitting is symmetrical with respect to the one reported in Figure \ref{figure: res_net_send}, with the characteristics of the receiving region having only a tenth of the impact of the ones of the sending region, leading to the interpretation that the probability that a researcher has migrated from a region, is guided mostly by the sending region's characteristics, which have a similar impact with respect to the ones of the model presented in Section \ref{subsub: Network GAM final model}.

\begin{figure}
\centering
\caption{\label{figure: res_net_both} Smooth components of the network model using both regions' researchers population. University Score for sending (A) and receiving (B) region, TED for sending (C) and receiving (D) region and distance between the two regions in log scale (E).}
\makebox{\includegraphics[scale=0.45]{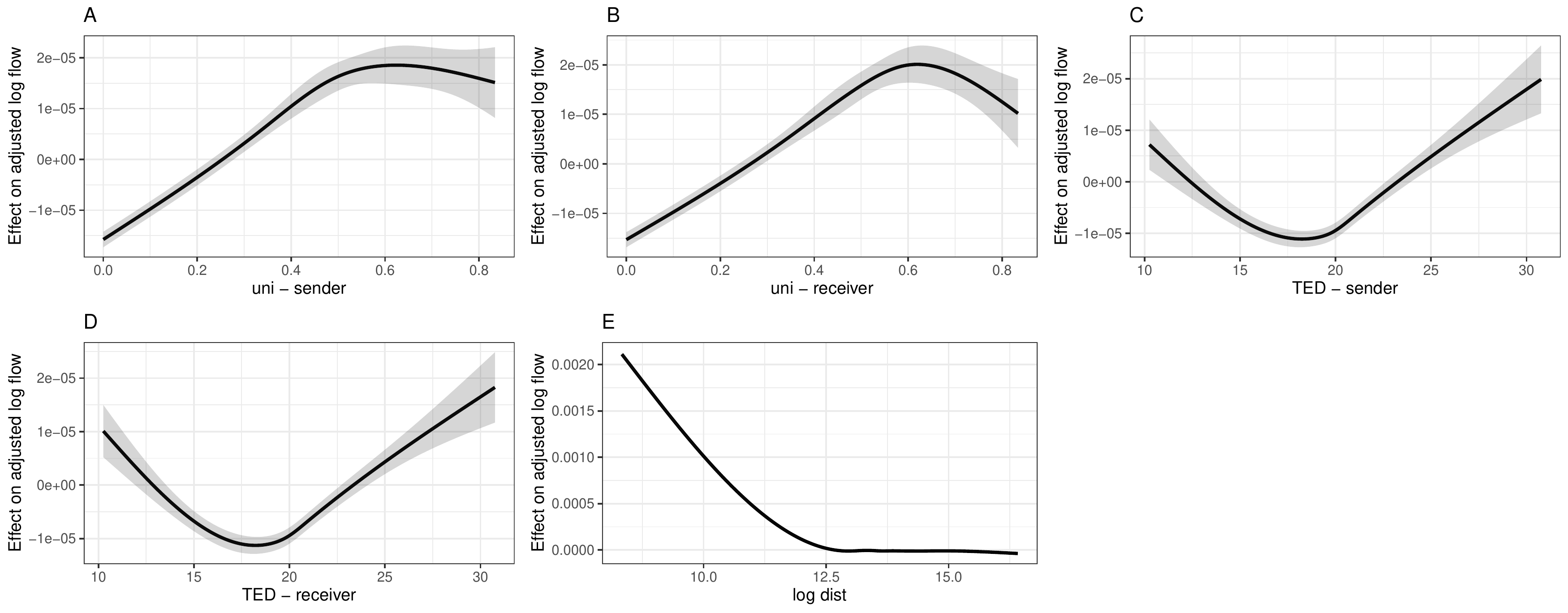}}
\end{figure}

In conclusion, when we consider the correction induced by both regions' research population size we are modeling the fraction of mobile researchers between the two regions across the whole research population. The visualization of its smoothed effects is reported in Figure \ref{figure: res_net_both}, where a strong similarity with the model proposed in Section \ref{subsub: Network GAM final model}.

\section{National Analysis}
\label{Appendix C: National Analysis}
National data are obtained directly from relative tables or as a composition of NUTS2 data reported in \ref{section:Data and preprocessing}. We hence obtain data for the 31 countries in the spatial domain on which we repeat the principal steps of the work.

\subsection{National Preliminary Analysis}
\label{Appendix C national preliminary analyisis}
At the national level too we work under the framework of magnitude differences between years, which are hence accounted for during the analysis. Edge betweeness community detection and infomap community detection both find a single group composed by all countries of the domain. This result is coherent with the NUTS2 analysis of Section \ref{subsection: Spatial Characterization}, where regions are grouped all together or at the national level, respectively under the two interpretations.

\subsection{National Mobility Characterization}
\label{Appendix C national mobility characterization}
We notice that the raw correlation between national in flow and out flow is of 99.35\%. Spearman correlation index between flows as functions of time for each country is of 99.05\% and the p-value of Spearman correlation test is 0. These results are in accord with the NUTS2 analysis of Section \ref{section: Mobility Analysis}. The proposed framework of brain mobility is consistent when applied to countries.

\subsection{National Network Nodes Analysis}
\label{Appendix C national network nodes}
\begin{figure}
\centering
\caption{\label{figure: nat hubs aut hist}Histograms of hubs (A) and authorities (B) scores on cumulative national network.}
\makebox{\includegraphics[scale=0.45]{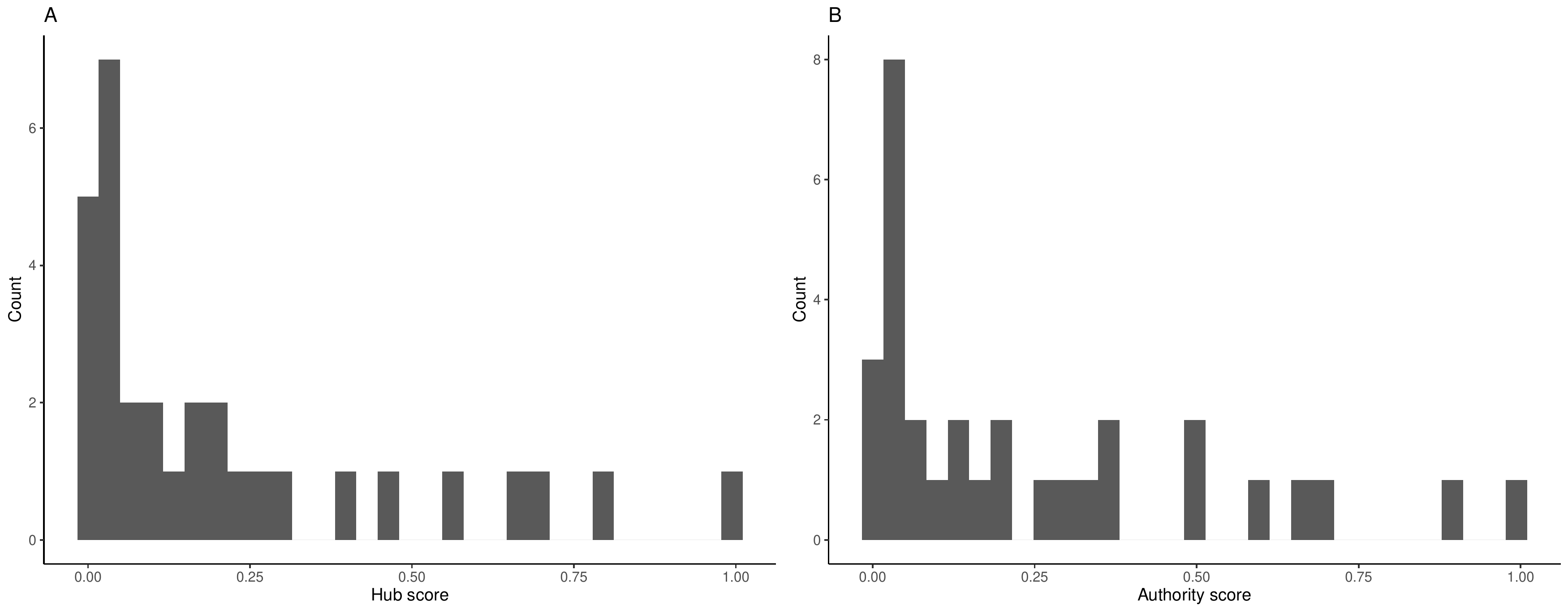}}

\end{figure}

We apply HITS algorithm on the cumulative national network, obtaining a correlation between the hubs and authorities score of 96.23\%. A visual representation of said scores is reported in Figure \ref{figure: nat hubs aut hist}. Given the shape of the distributions, we try and test a quantile based partition at 75\%, similarly to what was performed during the regional analysis. The countries selected as being with high scores, either hub or authority, are Switzerland, Germany, Spain, France, Italy, Netherlands, Portugal, Sweden and United Kingdom, while the remaining 22 countries are grouped together. University Score and TED result to be significantly different on such a partition, testing again with NPC-based ANOVA, results are reported in Table \ref{tab:ANOVA HA national}. We assess a general uniformity in time of this quantile partition on hubs and authorities score separately, computing them for each of 12 year-wise network and dividing the top 25\% countries from the others. The accuracy on groups belonging with respect to the cumulative network partition is between 0.93 and 1 for hubs, 0.87 and 1 for authorities. 

We propose the $s$-coreness analysis on the cumulative national network too, obtaining a 90\% correspondence between the top 25\% countries in term of $s$-core and the countries selected as having high hubs and authorities scores. 

Such results confirms that, just like in the regional scenario, countries which attract most are also the ones exporting most, moreover they represent the highly connected core of the network.

\begin{table}[t]
\begin{center}
\caption{\label{tab:ANOVA HA national}ANOVA test results for 75\% quantile hubs and authorities scores induced partition on national network.}
\begin{tabular}{@{}ccc@{}}
\toprule
Variable& Location p-value& Scale p-value\\
\midrule
GDP per capita & 0.334 & 0.498\\
Education Index & 0.088 & 0.465\\
University Score & 0.000 & 0.010\\
TED & 0.002 & 0.946\\
\bottomrule
\end{tabular}
\end{center}
\end{table}

\subsection{National Mobility Models}
\label{Appendix C national mobility model}
\begin{figure}
\centering
\caption{\label{national mobility model}National level mobility model's smooth components. University Score (A) and TED (B).}
\makebox{\includegraphics[scale=0.45]{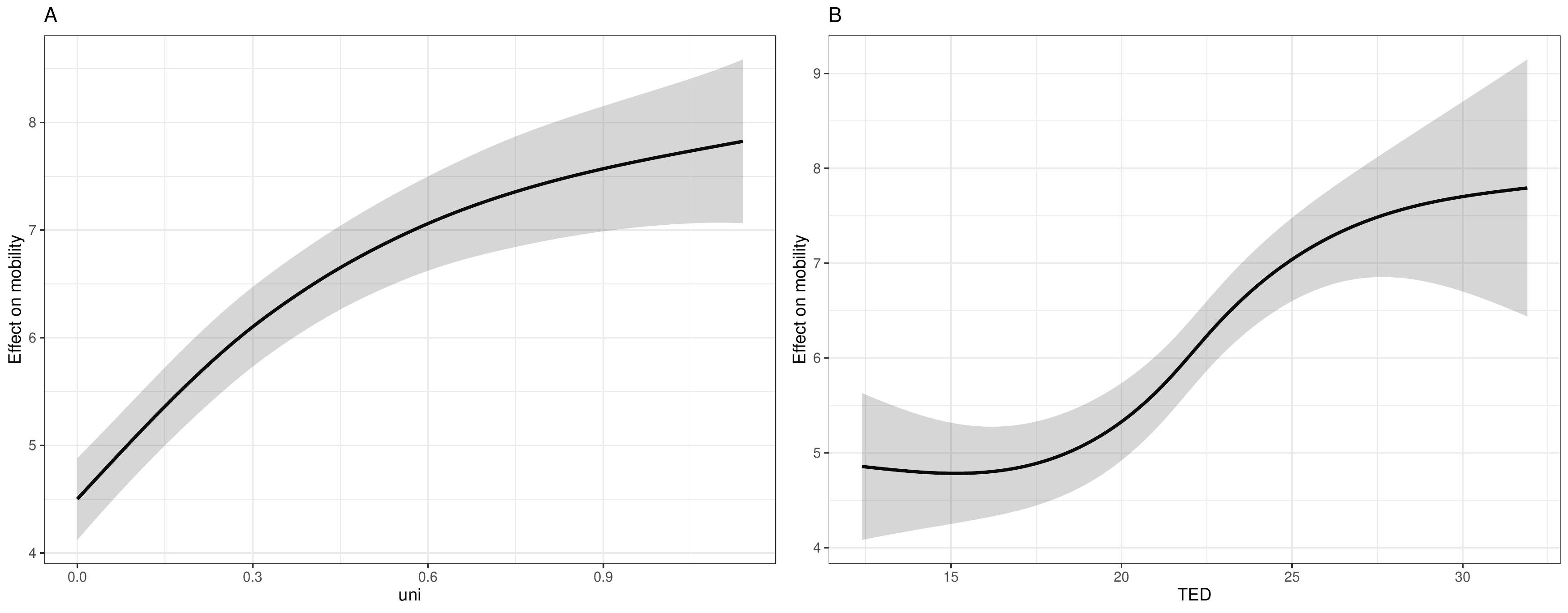}}
\end{figure}

We fit the Model \ref{national_mob} for the logarithm of the total flow as sum of incoming and outgoing flows for each country, using the same variables as for Model \ref{final mobility model}, obtaining n R-squared of 0.869. A visual representation of its smooth effects is reported in Figure \ref{national mobility model}. We notice that the relation between the response and the covariates strongly resembles the one of the mobility model in Section \ref{subsub: final mobility model}, therefore very similar conclusions can be drawn.

\begin{eqnarray}
\label{national_mob}
\log\left(in\_flow_{year}(country) +out\_flow_{year}(country) +1\right) &=& \overline{f}^{u}(uni_{year}(country))+\nonumber\\
&+& \overline{f}^{t}(TED_{year}(country))+\\
&+& \overline{\alpha}_{year} +
\overline{\varepsilon}_{year}\nonumber\\
\overline{\alpha}_{year} \overset{\textit{iid}}{\sim} \mathcal{N}(0, \overline{\sigma}_{year}), \ \overline{\varepsilon}_{year}^{region}\ \textit{iid}; && \forall\, country \in countries\ \forall\ years \in [2009, 2020]\nonumber
\end{eqnarray}

\subsection{National Network Models}
\label{Appendix C national network model}

\begin{figure}
\centering
\caption{\label{national network model}National level network model’s smooth components, University Score for sending (A) and receiving (B) region, TED for sending (C) and receiving (D)
region and distance between the two regions in log sc}
\makebox{\includegraphics[scale=0.45]{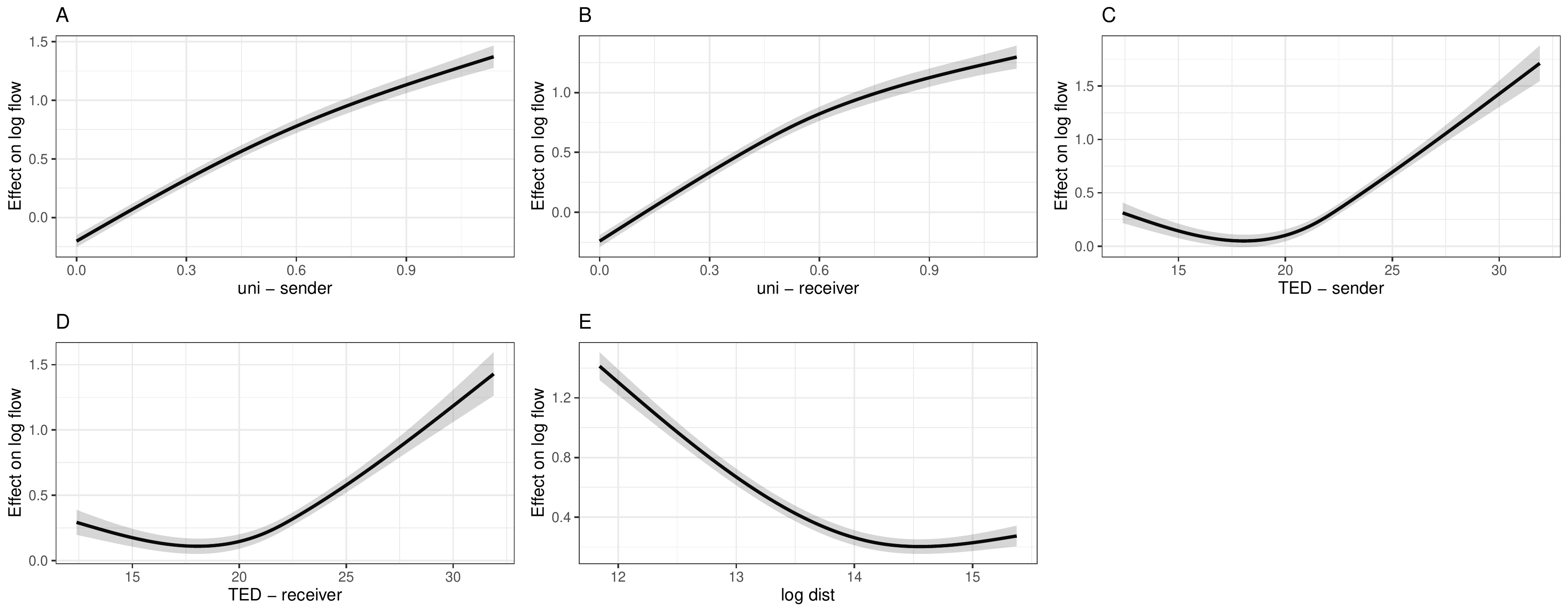}}
\end{figure}

We fit the Model \ref{national_network_model} for the logarithm of the flow from a country to another, on top of a directed weighted network of countries, using the same variables as in Model \ref{mod_network_final}. It achieves an R-squared of 0.795, a visual representation of its smooth effects is reported in Figure \ref{national network model}. 
We notice that the effects of sender and receiver countries strongly resemble the ones of the network model in Section \ref{subsub: Network GAM final model}, are again symmetrical and therefore similar conclusions can be drawn. 

\begin{eqnarray}\label{national_network_model}
\log\left(flow_{year}(sender,receiver)+1\right) &=&  \overline{f}^{u, s}(uni_{year}(sender)) +\overline{f}^{u, r}(uni_{year}(receiver)) +\nonumber \\
&+& \overline{f}^{t, s}(TED_{year}(sender)) +\overline{f}^{t, r}(TED_{year}(receiver)) +\nonumber \\
&+& \overline{g}(\log dist(sender, receiver)) + \\
&+& \overline{\alpha}_{year} +
\overline{\varepsilon}_{year}^{(sender, receiver)}\nonumber\\
 \overline{\alpha}_{year} \overset{\textit{iid}}{\sim} \mathcal{N}(0, \overline{\sigma}), \ \overline{\varepsilon}_{year}^{(sender, receiver)}\ \textit{iid}; &&\forall\, sender,\ receiver \in countries\, \forall\ year \in [2009, 2020]\nonumber
\end{eqnarray}

\section{Incoming and Outgoing Flows Analayses}

\subsection{Mobility Models for Incoming and Outgoing Flows} 
\label{Appendix D mob mod in out}
\begin{figure}
\centering
\caption{\label{in mod}Final ingoing flow model's smooth components, University Score (A), TED (B) and Education Index (C).}
\makebox{\includegraphics[scale=0.45]{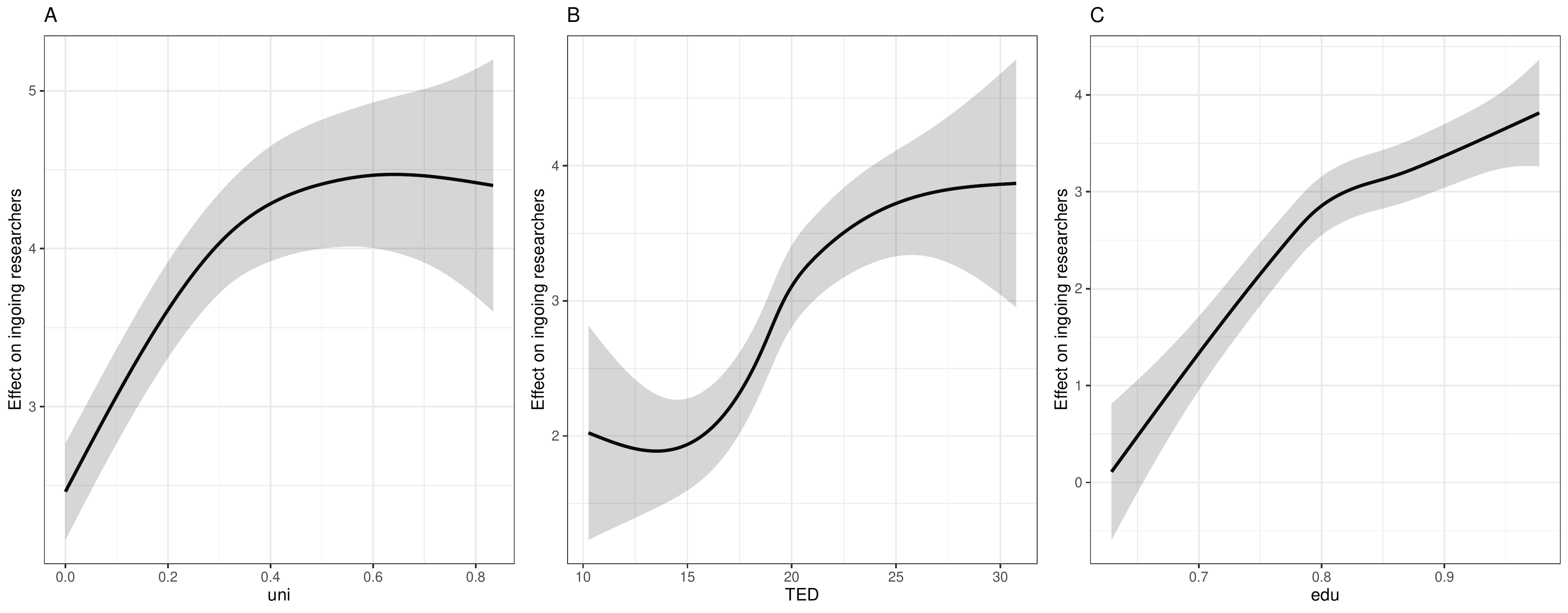}}
\end{figure}

\begin{figure}
\centering
\caption{\label{out mod}Final outgoing flow model's smooth components, University Score (A), TED (B) and Education Index (C).}
\makebox{\includegraphics[scale=0.45]{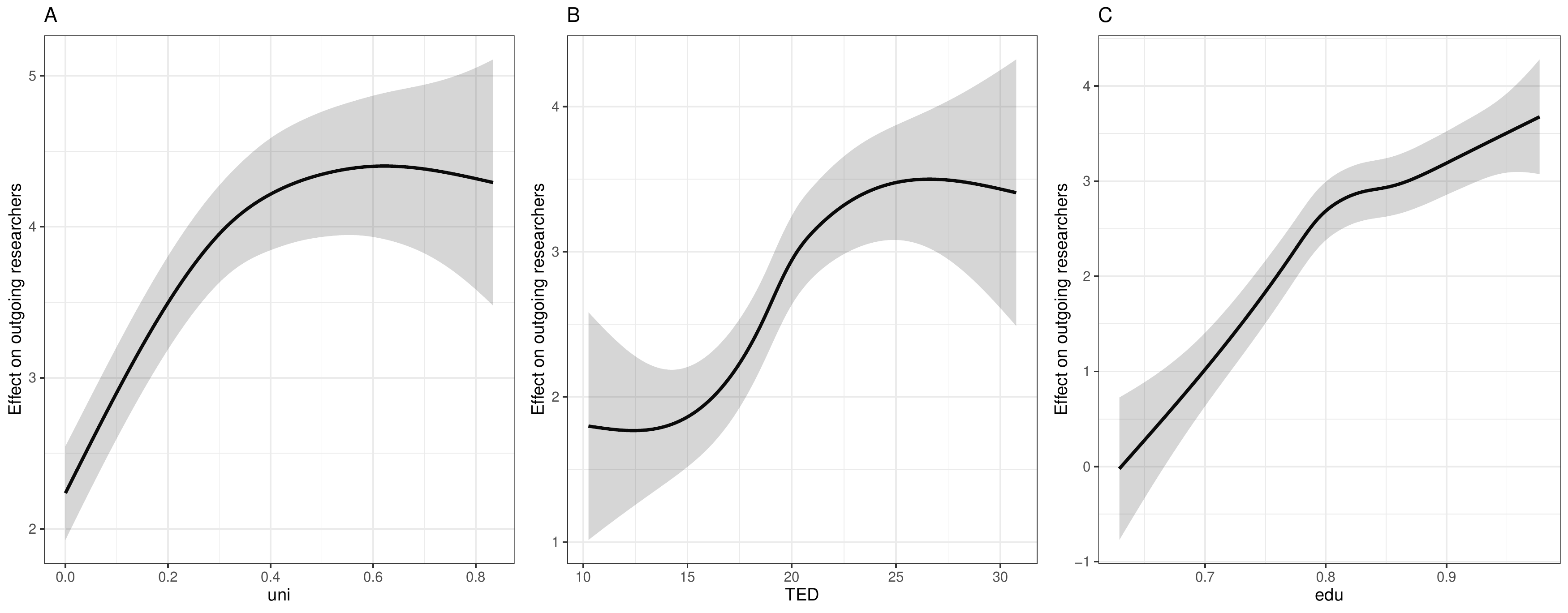}}
\end{figure}

We fit the mobility model where the response is either the total number of researchers going in a region or out of a region. As expected from the high correlation between the incoming and outgoing flow, the effects of the variables are extremely similar between the two models and with respect to the total flow model presented in Section \ref{subsection: Mobility Model}. In particular, the model for incoming flow obtains a R-squared of 0.679 and its smooth effects are reported in Figure \ref{in mod}, while the model for outgoing flow achieves a R-squared of 0.685 and its smooth effects are reported in Figure \ref{out mod}.

\subsection{PCA on Incoming and Outgoing Flows} 
\label{Appendix D pca in out}

Table \ref{tab:PCA out flow} and Table \ref{tab:PCA in flow} show the loadings for the first three principal components on data for outgoing and incoming flows respectively. Analogous considerations as for the total flow case on time heterogeneity can be made. These results are not surprising as the in flows and out flows of a region are strongly correlated. 

\begin{table}[t]
\begin{center}
\caption{\label{tab:PCA out flow}PCA summary for first 3 principal components on outgoing flows.}
\begin{tabular}{@{}c|*{3}{r}|*{3}{r}@{}}
\toprule
&\multicolumn{3}{c|}{ Total Flows}& \multicolumn{3}{c}{ Normalized Flows} \\
& \textbf{1st PC} & \textbf{2nd PC} & \textbf{3rd PC}& \textbf{1st PC} & \textbf{2nd PC} & \textbf{3rd PC}\\\hline\hline
 Proportion of variance explained & 0.3421 &  0.2519  & 0.1759   & 0.1995 & 0.1189 &   0.1006 \\\hline
Year &\multicolumn{6}{c}{ Component's loading}\\\hline
2009 &  &  &                   &  0.246 & 0.132 & 0.424  \\              
2010 &  &  &                   &  0.222 &-0.172 &-0.321   \\              
2011 & 0.101  &  &             &  0.357 & 0.464 &-0.214 \\              
2012 &  &  &                   &  0.200 &-0.309 &-0.111      \\             
2013 &  &  &                   &  0.317 &       & 0.531   \\              
2014 &  & 0.142  &             &  0.152 & 0.123 &-0.258  \\              
2015 &  &  & 0.168             &  0.414 &-0.338 &         \\          
2016 &  &  &                   &  0.276 & 0.128 & 0.261  \\              
2017 &  &  &                   &  0.342 &-0.145 & 0.272   \\              
2018 & 0.953 &-0.284 &         &  0.222 & 0.350 &-0.140 \\    
2019 &  0.256 & 0.926 &-0.196  &  0.257 & 0.398 &-0.286  \\ 
2020 &  & 0.169 & 0.952        &  0.345 &-0.440 &-0.241  \\
\bottomrule
\end{tabular}
\end{center}
\end{table}

\begin{table}[t]
\begin{center}
\caption{\label{tab:PCA in flow}PCA summary for first 3 principal components on ingoing flows.}
\begin{tabular}{@{}c|*{3}{r}|*{3}{r}@{}}
\toprule
&\multicolumn{3}{c|}{Total Flows}& \multicolumn{3}{c}{ Normalized Flows} \\
& \textbf{1st PC} & \textbf{2nd PC} & \textbf{3rd PC}& \textbf{1st PC} & \textbf{2nd PC} & \textbf{3rd PC}\\\hline\hline
Proportion of variance explained & 0.3327  &  0.2760  &  0.1858 & 0.1968 & 0.1154 & 0.09915 \\\hline
\em Year &\multicolumn{6}{c}{\em Component's loading}\\\hline
2009 &  &  &                     &0.287 & 0.166 & 0.225  \\              
2010 &  &  &                     &0.257 &-0.206 &-0.436  \\              
2011 &  &  &                     &0.336 & 0.461 &-0.138  \\              
2012 &  &  &                     &0.205 &-0.310 &        \\             
2013 &  &  &                     &0.341 &       & 0.406  \\              
2014 &  &  &                     &0.145 &       &-0.519  \\              
2015 &  &  & 0.15                &0.399 &-0.348 &        \\          
2016 &  &  &                     &0.289 & 0.164 & 0.305  \\              
2017 &  &  &                     &0.332 &-0.103 & 0.286  \\              
2018 & 0.921 & -0.374 &          &0.223 & 0.391 &-0.128  \\    
2019 & 0.353 & 0.913& -0.157     &0.219 & 0.319 &-0.319  \\ 
2020 &  & 0.12 & 0.964           &0.331 &-0.448 &-0.106  \\
\bottomrule
\end{tabular}
\end{center}
\end{table}

\subsection{$S$-Coreness for Incoming and Outgoing strengths} 
\label{Appendix D s-coreness in out}

\begin{figure}
\begin{minipage}[c]{0.45\linewidth}
\caption{\label{s-core distr in} In-cores thresholds distribution.}
\makebox{\includegraphics[scale=0.35]{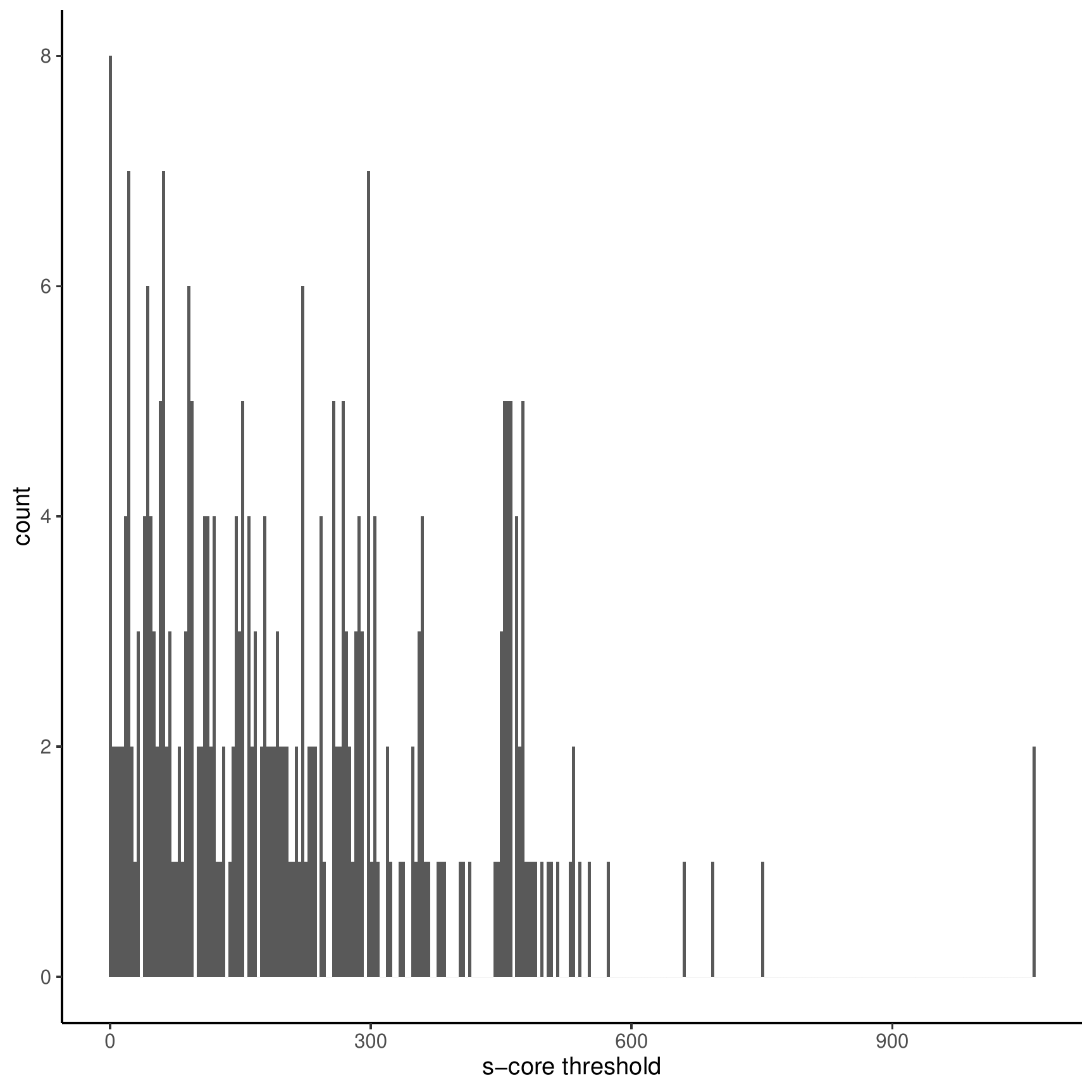}}
\end{minipage}
\hfill
\begin{minipage}[c]{0.45\linewidth}
\caption{\label{s-core distr out} Out-cores thresholds distribution.}
\makebox{\includegraphics[scale=0.35]{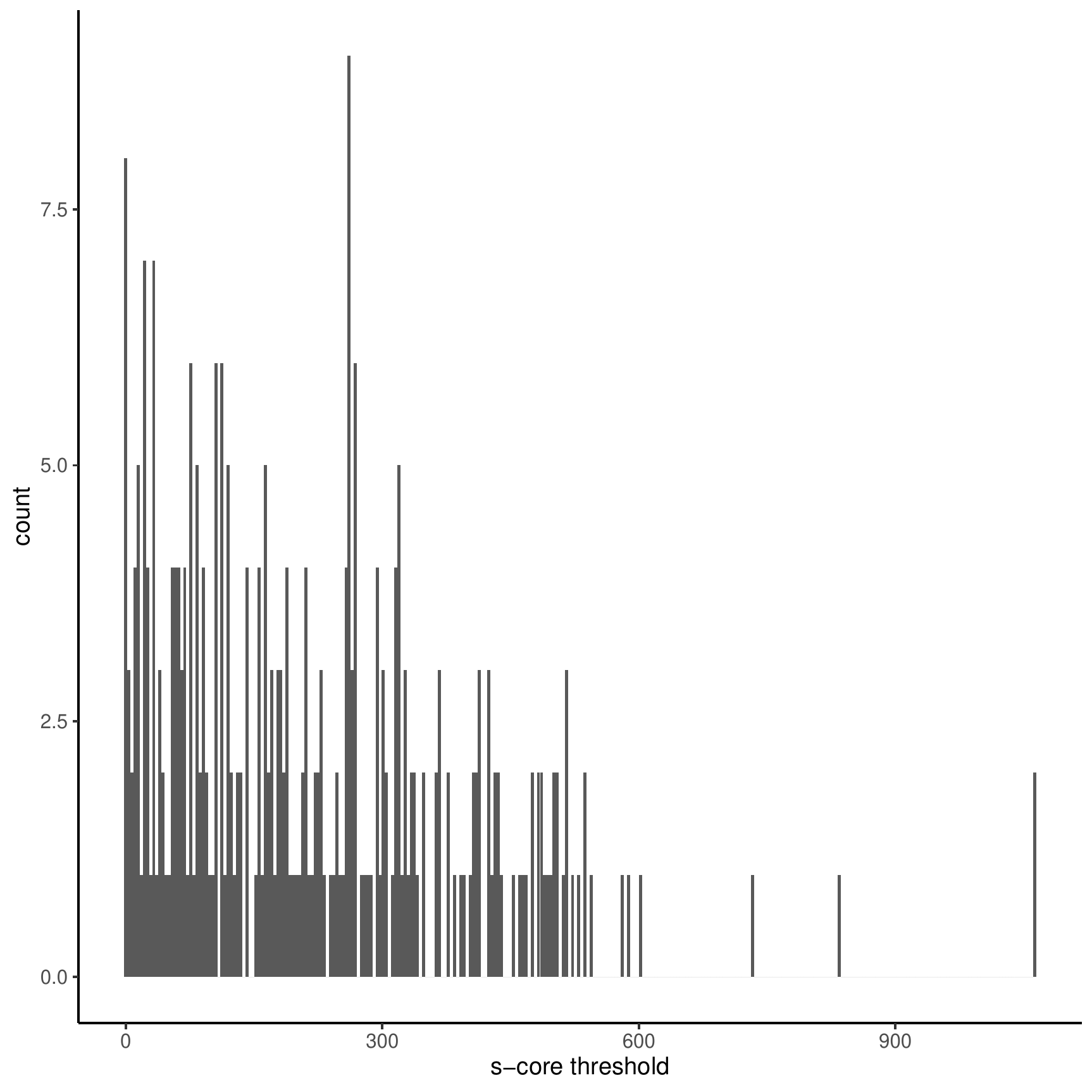}}
\end{minipage}%
\end{figure}

\begin{figure}
\begin{minipage}[c]{0.45\linewidth}
\caption{\label{shell in} In-core shell numerosity. }
\makebox{\includegraphics[scale=0.35]{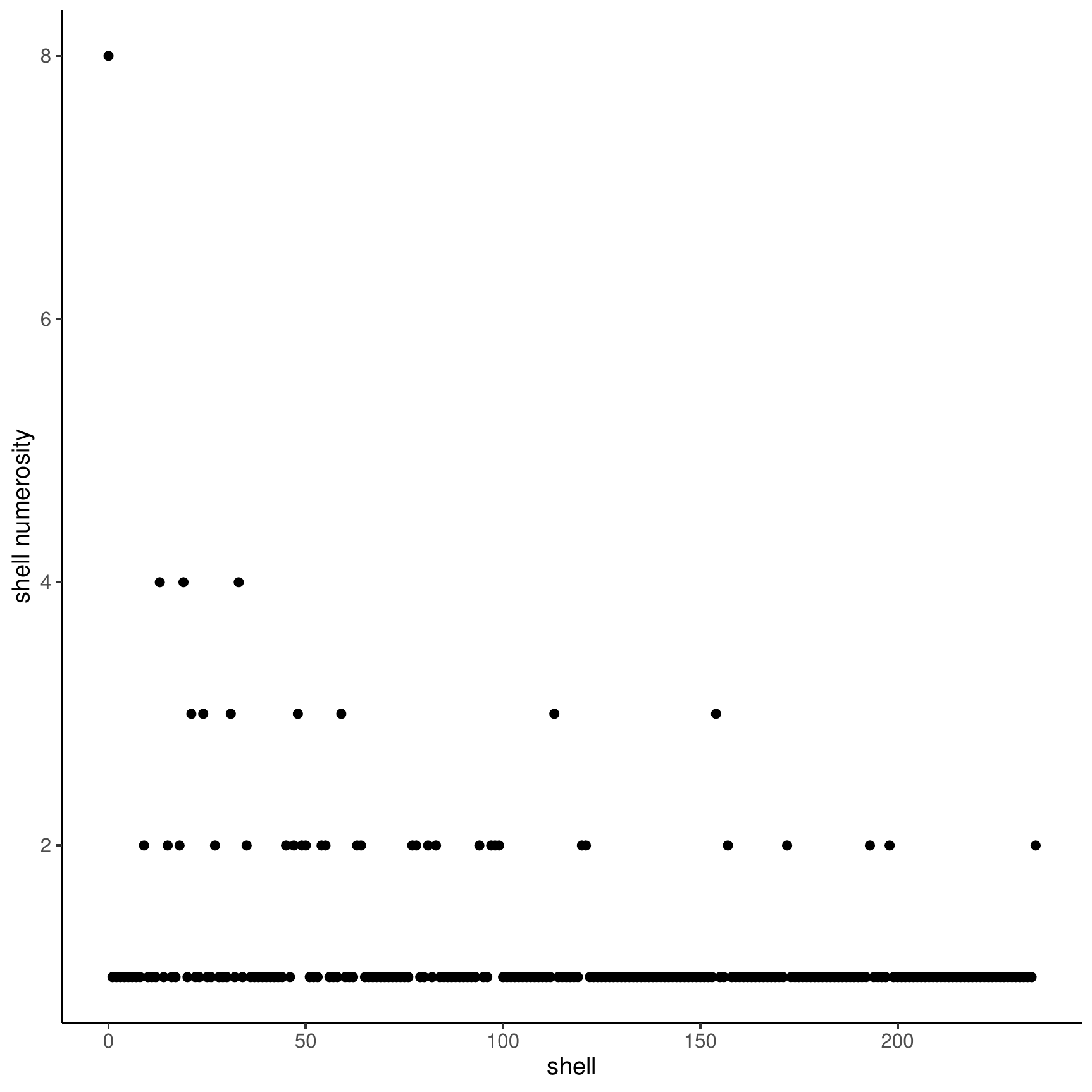}}
\end{minipage}
\hfill
\begin{minipage}[c]{0.45\linewidth}
\caption{\label{shell out} Out-core shell numerosity. }
\makebox{\includegraphics[scale=0.35]{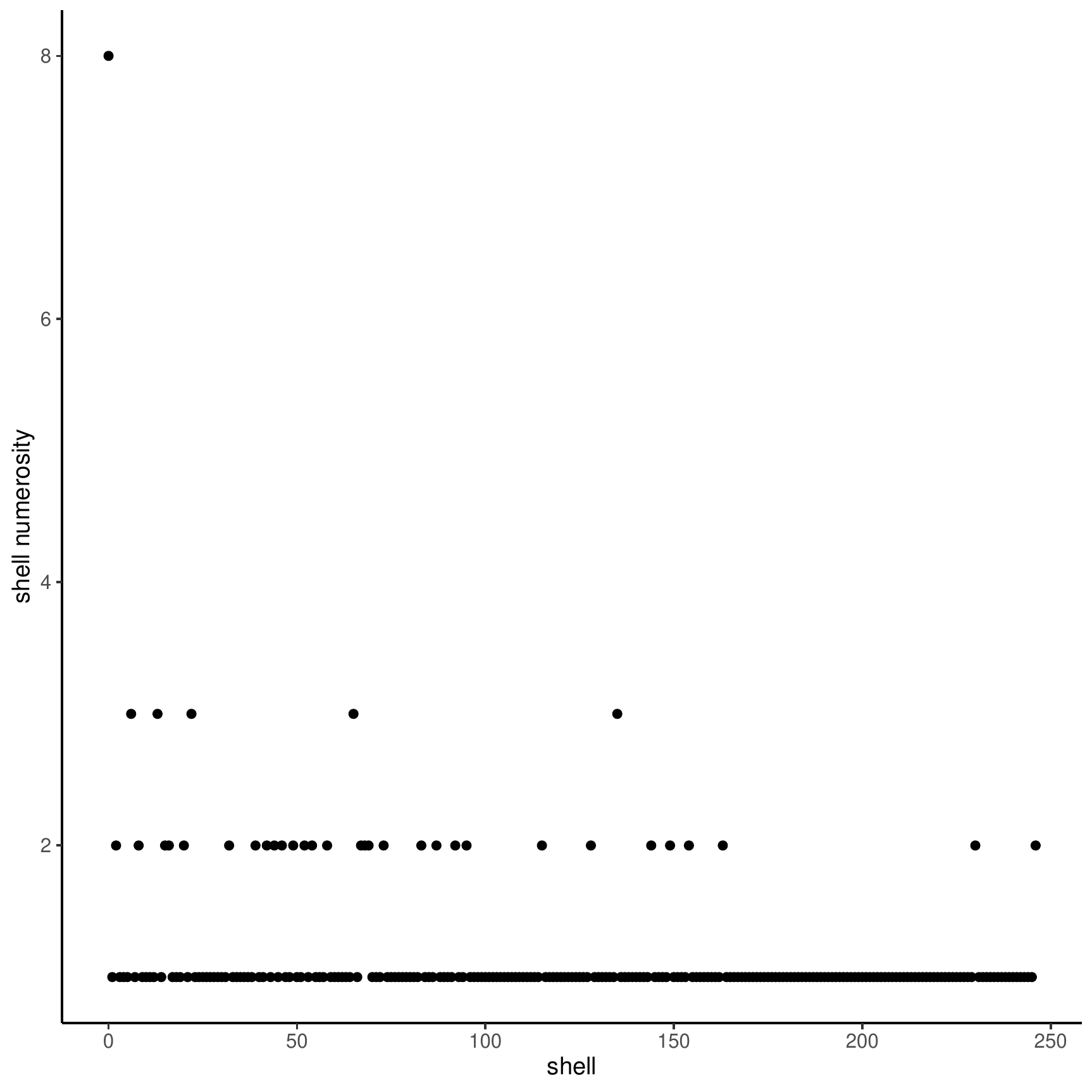}}
\end{minipage}%
\end{figure}

We perform $s$-coreness analysis for incoming and outgoing strength for the cumulative network, referring to them as in-score and out-score. 
We report distributions of the respective thresholds in Figure \ref{s-core distr in} and Figure \ref{s-core distr out} and shell numerosities in Figure \ref{shell in} and Figure \ref{shell out}, recognizing a strong similarity among them and with respect to thresholds distribution and shell numerosity of the total flow case inspected in Section \ref{subsub: S-coreness Analysis}.\\
We again notice a strong correspondence (71\%) between top 10\% hubs and authorities and top 10\% regions for $s$-coreness, both for in-score and out-score case.\\
We assess time omogeneity of the 90\% quantile partition for in-score and out-score, obtaining accuracies between 84\% and 90\% for the first, 85\% and 87\% for the second type of analysis.

%
%

\section{Hubs and Authorities Additional Analyses}
\label{appendix e hubs aut}

\subsection{Hubs and Authorities year-wise Partition}
\label{appendix e: hubs aut yearly anova}
Here we test, for each year included in the analysis, whether a partition on the top 10\% regions for hubs or authorities scores is reflected in the regional characteristics introduced in Section \ref{subsection:coovariates}, as supplement to the results of Section \ref{subsub: Hubs and Authorities Analysis}. A summary of the NPC-based ANOVA results is reported in Table \ref{tab: hubs aut yearly anova}. We can notice that University Score and TED are consistently significantly different between the groups identified by the partition induced by hubs and authorities score, year by year, while for the other variable, where the evidence for significant difference was weaker, if not even non-existent, the results are more heterogeneous. We hence deem the interpretations to be coherent with the ones obtained on the cumulative network.

\begin{table}[t]
\begin{center}

\caption{\label{tab: hubs aut yearly anova}ANOVA summary for year-wise partitions of regional characteristics.}
\begin{tabular}{@{}p{0.06\linewidth}|*{4}{p{0.06\linewidth}}|*{4}{p{0.06\linewidth}}@{}} 
\toprule
&\multicolumn{4}{c|}{ Location P-value}& \multicolumn{4}{c}{Scale P-value} \\ \hline
Year & Uni Score & Edu Index & GDP & TED & Uni Score & Edu Index & GDP & TED\\ \hline

2009 & 0 & 0.150 & 0.010 & 0 & 0 & 0.080 & 0.900 & 0.060  \\              
2010 & 0 & 0.670 & 0.670 & 0 & 0 & 0.540 & 0.510 & 0.900  \\              
2011 & 0 & 0.700 & 0.086 & 0 & 0 & 0.700 & 0.570 & 0.700   \\              
2012 & 0 & 0.700 & 0.086 & 0 & 0 & 0.700 & 0.570 & 0.700 \\             
2013 & 0 & 0 & 0.170 & 0 & 0 & 0.080 & 0.253 & 0.700  \\              
2014 & 0 & 0.006 & 0.059 & 0 & 0 & 0.860 & 0.130 & 0.700   \\              
2015 & 0 & 0 & 0.190 & 0 & 0 & 0.120 & 0.110 & 0.980   \\          
2016 & 0 & 0.009 & 0.420 & 0 & 0 & 0.153 & 0.200 & 0.700   \\              
2017 & 0 & 0.003 & 0.103 & 0 & 0 & 0.080 & 0.180 & 0.400  \\              
2018 & 0 & 0.001 & 0.135 & 0 & 0 & 0.050 & 0.570 & 0.800   \\    
2019 & 0 & 0.002 & 0.048 & 0 & 0 & 0.106 & 0.990 & 0.870  \\ 
2020 & 0 & 0.001 & 0.010 & 0 & 0 & 0.185 & 0.422 & 0.700  \\
\bottomrule
\end{tabular}
\end{center}
\end{table}

\section{Competing interests}
No competing interest is declared.

\section{Code availability}
The code used to produce the results of the analysis and the plots presented in this paper, together with the processed datasets, are reported in the GitHub repository \href{https://github.com/MartaMastropietro/Researchers-Migrations}{https://github.com/MartaMastropietro/Researchers-Migrations}.

\section{Author contributions statement}
J.G., M.M. and M.F. conceived the research, M.F. coordinated the research, J.G. and M.M. gathered and analysed the data, J.G., M.M., M.F., S.V. and F.I. commented the results, J.G., M.M. and M.F. wrote the manuscript, J.G., M.M., M.F., S.V. and F.I. reviewed the manuscript.


\cleardoublepage

\end{document}